\documentclass[12pt]{article}

\usepackage[cp1250]{inputenc}
\usepackage[verbose,a4paper,tmargin=0.5in,lmargin=1in,rmargin=1in]{geometry}
\usepackage[pdftex]{graphicx}
\usepackage{amsfonts}
\usepackage{indentfirst}
\usepackage{latexsym}
\usepackage{amsmath}
\usepackage{amsthm}
\usepackage{amssymb}
\usepackage{psfrag}
\usepackage[mathscr]{eucal} 
\usepackage{color}
\usepackage{cite}
\usepackage{longtable}
\usepackage{verbatim}
\usepackage{mathabx}

\usepackage{graphicx}
\usepackage[all]{xy}
\usepackage{float}

\usepackage{physics}
\usepackage{etoolbox}

\newfloat{Scheme}{tbp}{muj}

\newcommand\quotientfour[2]{{^{\displaystyle #1}}\Big/{_{\displaystyle #2}}}

\title{Curvature and conformal curvature dynamics formalisms and their applications in linearized gravity}

\author{$\textrm{Adam Chudecki}^{*}$ and $\textrm{Maciej Przanowski}^{**}$}

\begin{document}

\maketitle

$*$ Center of Mathematics and Physics, Lodz University of Technology, Al. Politechniki 11, 90-924 \L{}\'od\'z, Poland, adam.chudecki@p.lodz.pl
\newline
\newline
$**$ Institute of Physics, Lodz University of Technology, W\'olcza\'nska 219, 90-924 \L{}\'od\'z, Poland, maciej.przanowski@p.lodz.pl
\\
\\
\textbf{Abstract}. Tensorial, spinorial and helicity formalisms of the curvature and conformal curvature dynamics are developed. Equations of linearized gravity within that formalisms are given. Gravitational radiation in linearized gravity in terms of curvature dynamics is investigated. Equivalence of the Bia\l{}ynicki-Birula formula for the gravitational energy in linearized gravity and the Landau-Lifschitz formula is proved. Analogous result is found for the momentum in linearized gravity.


\section{Introduction}
\setcounter{equation}{0}

In this work we adopt and develop in some directions the idea presented by Jan W. van Holten in his distinguished paper \cite{b1} which can be summarized in his own words as: "However, the relative acceleration between local inertial frames at different points in space at different times is encoded in the space-time curvature and cannot be eliminated by any choice of reference frame. Therefore, an essential description of gravitational is to be cast in terms of the dynamics of curvature". To realize this programme, we deal with the general curvature and conformal curvature dynamics in tensorial, spinorial and helicity formalisms and then we employ the results obtained to the case of linearized theory of weak gravitational radiation. Although the most of results are well known from the papers and textbooks on general relativity the presentation of those results in terms of the curvature variables gives a new insight into theory of gravitation. To confirm the last statement it is worth quoting the words from an inspiring work by I. Bia\l{}ynicki-Birula \cite{b2}: "I have shown that the quantization of the linearized gravitational field that employs only the Riemann tensor (and not the metric tensor) can be achieved without any reference to the canonical formalism. In this approach the complications arising in the process of extracting true degrees of freedom never appear". 

Our paper is organized as follows. In section 2 we find the main equations describing the curvature and conformal curvature dynamics within the tensorial and spinorial formalisms. These equations are derived from the Bianchi identities and from the Einstein equations. In section 3 the same equations are investigated in terms of Pleba\'nski's helicity formalism \cite{b3} and the gravito-electromagnetism. The close analogy between gravity and electromagnetism is carefully studied in the helicity formalism language. The equations obtained in sections 2 and 3 are then specified to the case of linearized gravity in section 4. Section 5 is devoted to the theory of gravitational radiation in linearized gravity. We investigate this theory in terms of the curvature dynamics employing the electric and magnetic parts of free gravitational field considered in sections 3 and 4. In particular Eqs. (\ref{pomocnicze_41_a}), (\ref{pomocnicze_41_b}) and (\ref{postac_L}) give the power carried by the gravitational radiation in linearized gravity. Eq. (\ref{postac_L}) presents a well known result \cite{b4, b5, b6}. Finally, in section 6 we prove that the Bia\l{}ynicki-Birula formula for gravitational energy in vacuum in linearized gravity expressed purely with the use of electric and magnetic parts of free gravitational field \cite{b2} leads to the same result on gravitational energy as the Landau-Lifschitz pseudotensor and the Einstein pseudotensor if the appropriate factor $\dfrac{c^{4}}{64 \pi^{2}G}$ is assumed in that formula. Analogous considerations are done for the formula defining the momentum of gravitational field in terms of the electric and magnetic parts of this field.


\renewcommand{\arraystretch}{1.0}
\setlength\arraycolsep{2pt}
\setcounter{equation}{0}

\section{Curvature and Weyl curvature dynamics: tensorial and spinorial formalism}
\label{section_curvature}
\setcounter{equation}{0}

We deal with a spacetime equipped with the metric
\begin{equation}
ds^{2} = g_{\mu \nu} \, dx^{\mu} \otimes dx^{\nu}, \ \ g_{\mu \nu} = g_{\nu \mu}, \ \ \mu, \nu = 0,1,2,3
\end{equation}
of signature $(-+++)$. The \textsl{Levi-Civita connection coefficients read}
\begin{equation}
\label{Christofel_symbols}
\Gamma^{\mu}_{\  \nu \varrho} = \Gamma^{\mu}_{\  \varrho \nu} = \frac{1}{2} \, g^{\mu \tau} (\partial_{\nu} g_{\varrho \tau} + \partial_{\varrho} g_{\nu \tau} - \partial_{\tau} g_{\nu \varrho})
\end{equation}
where $g^{\mu \tau}$ is, as usually, the tensor inverse to $g_{\mu \tau}$
\begin{equation}
g_{\mu \nu} g^{\nu \varrho} = \delta^{\varrho}_{\mu}
\end{equation}
and $\partial_{\nu} := \dfrac{\partial}{\partial x^{\nu}}$. For a definition of the \textsl{curvature tensor} $R^{\alpha}_{\  \beta \gamma \delta}$ we assume the convention
\begin{equation}
R^{\alpha}_{\  \beta \gamma \delta} = \partial_{\gamma} \Gamma^{\alpha}_{\  \beta \delta} - \partial_{\delta} \Gamma^{\alpha}_{\  \beta \gamma} + \Gamma^{\alpha}_{\  \mu \gamma} \Gamma^{\mu}_{\  \beta \delta} - \Gamma^{\alpha}_{\  \mu \delta} \Gamma^{\mu}_{\  \beta \gamma}
\end{equation}
Under such a convention the commutator of covariant derivatives reads
\begin{equation}
\label{komutator_pochodnych_kowariantnych}
[\nabla_{\gamma}, \nabla_{\delta}] v^{\alpha} = v^{\beta} R^{\alpha}_{\  \beta \gamma \delta}
\end{equation}
Then the \textsl{Ricci tensor} and the \textsl{curvature scalar} are given as
\begin{equation}
\label{Ricci_tensor}
R_{\beta \gamma} =  R^{\alpha}_{\  \beta \gamma \alpha}, \ \ R_{\beta \gamma} = R_{\gamma \beta}
\end{equation}
and 
\begin{equation}
R = R^{\beta}_{\ \beta}
\end{equation}
respectively. The \textsl{Riemann curvature tensor} is defined as
\begin{eqnarray}
\label{Riemann_curvature_tensor_definition}
R_{\alpha \beta \gamma \delta} = g_{\alpha \mu} R^{\mu}_{\ \beta \gamma \delta} &=& \frac{1}{2} \, (\partial_{\beta} \partial_{\gamma} g_{\alpha \delta} + \partial_{\alpha} \partial_{\delta} g_{\beta \gamma} - \partial_{\alpha} \partial_{\gamma} g_{\beta \delta} - \partial_{\beta} \partial_{\delta} g_{\alpha \gamma})
\\ \nonumber
&&+ g_{\mu \nu} (\Gamma^{\mu}_{\ \beta \gamma} \Gamma^{\nu}_{\ \alpha \delta} - \Gamma^{\mu}_{\ \beta \delta} \Gamma^{\nu}_{\ \alpha \gamma})
\end{eqnarray}
Then the \textsl{Weyl tensor (Weyl curvature tensor, the Weyl conformal tensor)} is given by the formula
\begin{equation}
\label{Weyl_tensor_definition}
C_{\alpha \beta \gamma \delta} = R_{\alpha \beta \gamma \delta} + \frac{1}{2} \, (g_{\alpha \gamma} C_{\beta \delta} + g_{\beta \delta} C_{\alpha \gamma} - g_{\beta \gamma} C_{\alpha \delta} - g_{\alpha \delta} C_{\beta \gamma}) + \frac{R}{12} \, (g_{\alpha \gamma} g_{\beta \delta} - g_{\alpha \delta} g_{\beta \gamma})
\end{equation}
where 
\begin{equation}
\label{tracelss_Ricci}
C_{\alpha \beta} = R_{\alpha \beta} - \frac{1}{4} g_{\alpha \beta} R
\end{equation}
stands for the \textsl{traceless Ricci tensor}
\begin{equation}
C^{\alpha}_{\ \alpha}=0
\end{equation}
With the use of (\ref{tracelss_Ricci}) the Einstein equations
\begin{equation}
\label{Rownania_Einsteina}
R_{\alpha \beta} - \frac{1}{2} R g_{\alpha \beta} = - \frac{8 \pi G}{c^{4}} \, T_{\alpha \beta}
\end{equation}
where $T_{\alpha \beta}$ is the energy-momentum tensor ($G$ is the gravitational constant and $c$ denotes the speed of light) can be rewritten as 
\begin{subequations}
\begin{eqnarray}
\label{Rownania_Einsteina_second_version}
C_{\alpha \beta} &=& - \frac{8 \pi G}{c^{4}} \left( T_{\alpha \beta} - \frac{1}{4} \, T g_{\alpha \beta} \right)
\\
\label{matter_skalar}
R &=& \frac{8 \pi G}{c^{4}} T, \ \ T = T^{\alpha}_{\ \alpha}
\end{eqnarray}
\end{subequations}
The crucial role in the curvature and Weyl curvature dynamics is played by the Bianchi identities. In standard form they read
\begin{equation}
\label{Bianchi_identities}
R^{\alpha}_{\ \beta [ \gamma \delta ; \varrho]} = 0
\end{equation}
where the square bracket $[...]$ stands for the anti-symmetrization and the symbol "$_{; \varrho}$" means the covariant derivative, $_{; \varrho} \equiv \nabla_{\varrho}$. Contraction with respect to $\alpha$ and $\varrho$ gives (see (\ref{Ricci_tensor}))
\begin{equation}
\label{Bianchi_identities_contraction}
R^{\alpha}_{\ \beta \gamma \delta ; \alpha} = 2 R_{\beta [ \gamma ; \delta]}
\end{equation}
One can show that the identities (\ref{Bianchi_identities_contraction}) are equivalent to (\ref{Bianchi_identities}). Employing the Einstein field equations (\ref{Rownania_Einsteina}) and (\ref{matter_skalar}) we write (\ref{Bianchi_identities_contraction}) in the form
\begin{eqnarray}
\label{pomocnicze_3}
R^{\alpha}_{\ \beta \gamma \delta ; \alpha} = - \frac{16 \pi G}{c^{4}} \, \widecheck{T}_{\beta [\gamma ; \delta]}
\\ \nonumber
\widecheck{T}_{\beta \gamma} = T_{\beta \gamma} - \frac{1}{2} T g_{\beta \gamma}
\end{eqnarray}
Equivalently, using (\ref{Weyl_tensor_definition}) and (\ref{tracelss_Ricci}) one can rewrite the Bianchi identities (\ref{Bianchi_identities_contraction}) in terms of the Weyl tensor as \cite{b7, b8, b9}
\begin{equation}
\label{pomocnicze_1}
C^{\alpha}_{\ \beta \gamma \delta ; \alpha} = C_{\beta [\gamma ; \delta ]} + \frac{1}{12} \, g_{\beta [ \gamma} R_{; \delta ]}
\end{equation}
or
\begin{subequations}
\begin{eqnarray}
\label{BI_1}
C^{\alpha}_{\ \beta \gamma \delta ; \alpha} &=& C_{\beta [ \gamma ; \delta ]} + \frac{1}{3} g_{\beta [ \gamma} C^{\varrho}_{\ \delta ] ; \varrho}
\\ 
\label{BI_2}
C^{\alpha}_{\ \beta ; \alpha} &=& \frac{1}{4} R_{; \beta}
\end{eqnarray}
\end{subequations}
Inserting (\ref{Rownania_Einsteina_second_version}) and (\ref{matter_skalar}) into (\ref{pomocnicze_1}) we obtain
\begin{equation}
\label{pomocnicze_2}
C^{\alpha}_{\ \beta \gamma \delta ; \alpha} = - \frac{8 \pi G}{c^{4}} \left( T_{\beta [ \gamma ; \delta ]} - \frac{1}{3} g_{\beta [ \gamma} T_{; \delta]} \right)
\end{equation}
Using in (\ref{pomocnicze_2}) the well known relation
\begin{equation}
C^{\alpha \beta}_{\ \ \; \beta \delta} =0
\end{equation}
one quickly gets the matter equations
\begin{equation}
\label{matter_equation}
T^{\beta}_{\ \delta ; \beta} = 0
\end{equation}
Acting on (\ref{Bianchi_identities}) with the operator $g^{\sigma \varrho} \nabla_{\varrho}$ we get
\begin{eqnarray}
\label{pochodne_Bianchi_identities}
&&\nabla^{\varrho} \nabla_{\varrho} R^{\alpha}_{\ \beta \gamma \delta} + \nabla_{\varrho} \nabla_{\delta} R^{\alpha \ \; \varrho}_{\ \beta \ \; \gamma} + \nabla_{\varrho} \nabla_{\gamma} R^{\alpha \ \ \varrho}_{\ \beta \delta} 
\\ \nonumber
&& = \nabla^{\varrho} \nabla_{\varrho} R^{\alpha}_{\ \beta \gamma \delta} + \nabla_{\varrho} \nabla_{\delta} R^{\varrho \ \; \alpha}_{\ \gamma \ \beta} - \nabla_{\varrho} \nabla_{\gamma} R^{\varrho \ \alpha}_{\ \delta \ \beta} = 0
\end{eqnarray}
Employing the general formulae of the commutators $[\nabla_{\varrho}, \nabla_{\delta}] R^{\varrho \ \alpha}_{\ \gamma \ \beta}$ and $[\nabla_{\varrho}, \nabla_{\gamma}] R^{\varrho \ \alpha}_{\ \delta \ \beta}$ which follow from (\ref{komutator_pochodnych_kowariantnych}) and then using (\ref{pomocnicze_3}), Einstein's equations (\ref{Rownania_Einsteina}) and algebraic properties of the curvature tensor we rewrite (\ref{pochodne_Bianchi_identities}) as
\begin{eqnarray}
\label{pomocnicze_4}
\nabla^{\varrho} \nabla_{\varrho} R_{\alpha \beta \gamma \delta} &=& \frac{16 \pi G}{c^{4}} \left( \widecheck{T}_{\gamma [ \alpha ; \beta] \delta} - \widecheck{T}_{\delta [\alpha ; \beta ] \gamma} + \widecheck{T}_{\mu [ \gamma} R^{\mu}_{\ \delta ] \alpha \beta} \right) 
\\ \nonumber
&& + 2R_{\alpha \mu \varrho \delta} R_{\beta \ \ \gamma}^{\ \, \mu \varrho} - 2 R_{\alpha \mu \varrho \gamma} R_{\beta \ \ \delta}^{\ \, \mu \varrho} - R_{\alpha \beta \mu \varrho} R^{\mu \varrho}_{\ \ \, \gamma \delta}
\end{eqnarray}
Taking into account that $R_{\alpha \beta \gamma \delta} = R_{\gamma \delta \alpha \beta}$ one can write (\ref{pomocnicze_4}) in another form \cite{b1}
\begin{eqnarray}
\label{drugie_pochodne_Bianchi_identities}
\nabla^{\varrho} \nabla_{\varrho} R_{\alpha \beta \gamma \delta} &=& \frac{8 \pi G}{c^{4}} \left( \widecheck{T}_{\alpha \gamma ; (\beta \delta)} + \widecheck{T}_{\beta \delta ; (\alpha \gamma)} - \widecheck{T}_{\alpha \delta; (\beta \gamma)} - \widecheck{T}_{\beta \gamma ; (\alpha \delta)} \right.
\\ \nonumber
&&
\ \ \ \ \ \ \ \ \ \ \left.
 + \widecheck{T}_{\mu [ \alpha} R^{\mu}_{\ \beta ] \gamma \delta} + \widecheck{T}_{\mu [\gamma} R^{\mu}_{\ \delta ] \alpha \beta} \right)
\\ \nonumber
&& + 2R_{\alpha \mu \varrho \delta} R_{\beta \ \ \gamma}^{\ \, \mu \varrho} - 2 R_{\alpha \mu \varrho \gamma} R_{\beta \ \ \delta}^{\ \, \mu \varrho} - R_{\alpha \beta \mu \varrho} R^{\mu \varrho}_{\ \ \, \gamma \delta}
\end{eqnarray}
where the bracket $(...)$ means the symmetrization. Inserting into (\ref{drugie_pochodne_Bianchi_identities}) $R_{\alpha \beta \gamma \delta}$ calculated from (\ref{Weyl_tensor_definition}) we obtain the equation for $\nabla^{\varrho} \nabla_{\varrho} C_{\alpha \beta \gamma \delta}$ (this equation is rather involved (see \cite{b1}, Eq. (21)) and we are going to consider it a little further (see Eq. (\ref{pomocnicze_19}))).

The main object which enables to move to the spinorial formalism is the \textsl{spinorial 1-form (the soldering form)} \cite{b10, b11, b12, b13}
\begin{equation}
g^{A \dot{B}} = g^{A \dot{B}}_{\ \ \ \mu} \, dx^{\mu}, \ \ \   A=1,2, \ \dot{B} = \dot{1}, \dot{2}
\end{equation}
We assume that in locally Galilean coordinate system the matrices $(g^{A \dot{B}}_{\ \ \ \mu})$, $\mu = 0,1,2,3$ take the form
\begin{equation}
\label{definicja_g_mu_AB}
(g^{A \dot{B}}_{\ \ \ \mu}) = \Bigg[
\left(\begin{array}{cc}
-1 & 0 \\
\ \ 0 & -1
\end{array}\right),
\left(\begin{array}{cc}
0 & \ \ 1 \\
1 & \ \ 0
\end{array}\right),
\left(\begin{array}{cc}
0 & -i \\
i & \ \  0
\end{array}\right),
\left(\begin{array}{cc}
1 & \ \ 0 \\
0 & -1
\end{array}\right) \Bigg]
\end{equation}
The tensorial indices are to be manipulated with the use of $g_{\mu \nu}$ or $g^{\mu \nu}$ and the spinorial indices according to the rules
\begin{eqnarray}
\label{spinorowe_podnoszenie_wskaznikow}
\chi^{A} = \in^{BA} \chi_{B}, \ \chi_{A} = \in_{AB} \chi^{B}, \ A,B = 1,2
\\ \nonumber
\chi^{\dot{A}} = \in^{\dot{B}\dot{A}} \chi_{\dot{B}}, \ \chi_{\dot{A}} = \in_{\dot{A}\dot{B}} \chi^{\dot{B}}, \ \dot{A}, \dot{B} = \dot{1},  \dot{2}
\end{eqnarray}
where the antisymmetric spinors $\in_{AB}$, $\in_{\dot{A}\dot{B}}$, $\in^{AB}$ and $\in^{\dot{A}\dot{B}}$ are defined as 
\begin{equation}
(\in_{AB}) = (\in^{AB}) = (\in_{\dot{A}\dot{B}}) = (\in^{\dot{A}\dot{B}}) = \left( \begin{array}{cc}
0 & \ \ 1 \\
-1 & \ \ 0
\end{array}\right)
\end{equation}
Recall that transformation of spinorial objects is given by 
\begin{eqnarray}
&& \Psi'^{A \dot{A}}_{\ \ \ B \dot{B}} = l^{A}_{\ E} \, l^{\dot{A}}_{\ \dot{E}} \, l^{-1 \, F}_{\ \ \ \ B} \, l^{-1 \, \dot{F}}_{\ \ \ \ \dot{B}} \, \Psi^{E \dot{E}}_{\ \ \ F \dot{F}}
\\ \nonumber
&& (l^{A}_{\ E}) \in SL(2; \mathbb{C}), \ l^{\dot{A}}_{\ \dot{F}} = \overline{l^{A}_{\ F}}, \ l^{-1 \, \dot{F}}_{\ \ \ \ \dot{B}} = \overline{l^{-1 \, F}_{\ \ \ \ B}}, \ l^{-1 \, F}_{\ \ \ \ B} \, l^{B}_{\ E} = \delta^{F}_{E}
\end{eqnarray}
where overbar denotes the complex conjugation. Define the following spinorial 2-forms
\begin{eqnarray}
\label{definicja_SAB}
&& S^{AB} = \frac{1}{2} S^{AB}_{\ \ \ \mu \nu} \, dx^{\mu} \wedge dx^{\nu} = \frac{1}{2} \in_{\dot{A}\dot{B}} \, g^{A \dot{A}} \wedge g^{B \dot{B}} = S^{(AB)}
\\ \nonumber
&& S^{\dot{A}\dot{B}} = \overline{S^{AB}} = \frac{1}{2} S^{\dot{A}\dot{B}}_{\ \ \ \mu \nu} \, dx^{\mu} \wedge dx^{\nu} = \frac{1}{2} \in_{AB} \, g^{A \dot{A}} \wedge g^{B \dot{B}} = S^{(\dot{A} \dot{B})}
\end{eqnarray}
With the use of these objects one defines the spinorial images of the Weyl tensor as
\begin{eqnarray}
\label{spinorial_image_Weyl}
C_{ABCD} &=& \frac{1}{16} \, S_{AB}^{\ \ \ \alpha \beta} \, S_{CD}^{\ \ \ \gamma \delta} \, C_{\alpha \beta \gamma \delta}
\\ \nonumber
C_{\dot{A} \dot{B}\dot{C}\dot{D}} &=& \overline{C_{ABCD}} = \frac{1}{16} \, S_{\dot{A}\dot{B}}^{\ \ \ \alpha \beta} \, S_{\dot{C}\dot{D}}^{\ \ \ \gamma \delta} \, C_{\alpha \beta \gamma \delta}
\end{eqnarray}
We can easily show that the \textsl{Weyl spinors} $C_{ABCD}$ and $C_{\dot{A}\dot{B}\dot{C}\dot{D}}$ are totally symmetric
\begin{equation}
C_{ABCD}=C_{(ABCD)}, \ C_{\dot{A}\dot{B}\dot{C}\dot{D}} = C_{(\dot{A}\dot{B}\dot{C}\dot{D})}
\end{equation}
The inverse relation to (\ref{spinorial_image_Weyl}) reads
\begin{equation}
\label{pomocnicze_15}
C_{\alpha \beta \gamma \delta} = \frac{1}{4} S^{AB}_{\ \ \ \alpha \beta} \, S^{CD}_{\ \ \ \gamma \delta} \, C_{ABCD} + \frac{1}{4} S^{\dot{A} \dot{B}}_{\ \ \ \alpha \beta} \, S^{\dot{C}\dot{D}}_{\ \ \ \gamma \delta} \, C_{\dot{A}\dot{B}\dot{C}\dot{D}}
\end{equation}
Then one can show the useful relation
\begin{equation}
\label{pomocnicze_15_a}
\ast C_{\alpha \beta \gamma \delta} = \frac{1}{4} S^{AB}_{\ \ \ \alpha \beta} \, S^{CD}_{\ \ \ \gamma \delta} \, C_{ABCD} - \frac{1}{4} S^{\dot{A} \dot{B}}_{\ \ \ \alpha \beta} \, S^{\dot{C}\dot{D}}_{\ \ \ \gamma \delta} \, C_{\dot{A}\dot{B}\dot{C}\dot{D}}
\end{equation}
where the \textsl{Hodge $\ast$-operation} is defined by
\begin{equation}
\ast C_{\alpha \beta \gamma \delta} = \frac{1}{2} i \sqrt{|\det (g_{\mu \nu})|} \, \in_{\alpha \beta \varrho \sigma} C_{\gamma \delta}^{\ \ \varrho \sigma}
\end{equation}
Bianchi identities in spinorial formalism read \cite{b9, b12, b13}
\begin{subequations}
\begin{eqnarray}
\label{Bianchi_identities_spinorial_1}
&& \nabla^{D \dot{D}} C_{ABCD} + \nabla_{(A}^{\ \ \dot{E}} C_{BC) \ \dot{E}}^{\ \ \ \; \dot{D}} = 0
\\
\label{Bianchi_identities_spinorial_2}
&& \nabla^{D\dot{D}} C_{\dot{A}\dot{B}\dot{C}\dot{D}} + \nabla^{E}_{\ (\dot{A}} C^{D}_{\ \; |E|\dot{B}\dot{C})}=0
\\
\label{Bianchi_identities_spinorial_3}
&& \nabla^{B \dot{D}} C_{AB \dot{C}\dot{D}} + \frac{1}{8} \nabla_{A \dot{C}} R=0
\end{eqnarray}
\end{subequations}
where 
\begin{equation}
\label{Ricci_traceless_spinorial}
C_{AB \dot{C}\dot{D}} = \frac{1}{4} \, g^{\ \ \ \; \mu}_{A \dot{C}} \, g^{\ \ \ \; \nu}_{B \dot{D}} \, C_{\mu \nu} = C_{(AB)\dot{C}\dot{D}} = C_{AB (\dot{C}\dot{D})}
\end{equation}
is the spinorial image of the traceless Ricci tensor $C_{\mu \nu}$ and
\begin{equation}
\label{nabla_spinorial}
\nabla^{A\dot{B}} = g^{A \dot{B} \mu} \, \nabla_{\mu}
\end{equation}
One can easily prove that the identities (\ref{Bianchi_identities_spinorial_1}) or their complex conjugate (\ref{Bianchi_identities_spinorial_2}) are equivalent to the identities (\ref{BI_1}), and the identities (\ref{Bianchi_identities_spinorial_3}) are equivalent to (\ref{BI_2}).

Inserting (\ref{Rownania_Einsteina_second_version}) into (\ref{Ricci_traceless_spinorial}) and using the relation
\begin{equation}
g^{\ \ \ \; \mu}_{A \dot{C}} \, g_{B \dot{D} \mu} = -2 \in_{AB} \in_{\dot{C}\dot{D}}
\end{equation}
we get
\begin{equation}
\label{pomocnicze_6}
C_{AB \dot{C}\dot{D}} = -\frac{8 \pi G}{c^{4}} \left( T_{AB \dot{C}\dot{D}} + \frac{1}{8} \in_{AB} \in_{\dot{C} \dot{D}} T \right)
\end{equation}
where
\begin{equation}
\label{tensor_matter_spinorial_image}
T_{AB \dot{C}\dot{D}} = \frac{1}{4}  \, g^{\ \ \ \; \mu}_{A \dot{C}}  \, g^{\ \ \ \; \nu}_{B \dot{D}}  \, T_{\mu \nu}
\end{equation}
is the spinorial image of the energy-momentum tensor $T_{\mu \nu}$. Substituting (\ref{pomocnicze_6}) into (\ref{Bianchi_identities_spinorial_1}) we obtain the equation
\begin{equation}
\label{pomocnicze_7}
\nabla^{D\dot{D}} C_{ABCD} = -\frac{8 \pi G}{c^{4}} \, \nabla_{(A |\dot{E}|} T_{BC)}^{\ \ \ \dot{D}\dot{E}}
\end{equation}
Analogously, substituting (\ref{pomocnicze_6}) into (\ref{Bianchi_identities_spinorial_2}) we obtain the complex conjugate of (\ref{pomocnicze_7})
\begin{equation}
\label{pomocnicze_7a}
\nabla^{D\dot{D}} C_{\dot{A}\dot{B}\dot{C}\dot{D}} = - \frac{8 \pi G}{c^{4}} \, \nabla_{E(\dot{A}} T^{DE}_{\ \ \ \; \dot{B}\dot{C})}
\end{equation}
Finally, from (\ref{pomocnicze_6}), (\ref{matter_skalar}) and (\ref{Bianchi_identities_spinorial_3}) one gets the matter equation (\ref{matter_equation}) in spinorial language
\begin{equation}
\nabla^{B \dot{D}} T_{AB \dot{C}\dot{D}}=0
\end{equation}
To proceed further we act on both sides of Eq. (\ref{pomocnicze_7}) with $\nabla_{F \dot{D}}$. Employing then the definition (\ref{nabla_spinorial}) and the formula \cite{b12}
\begin{eqnarray}
g^{\ \ \ \; \mu}_{F \dot{D}} g^{D \dot{D} \nu} &=& \frac{1}{2} (g^{D \dot{D} \nu} g^{\ \ \ \; \mu}_{F \dot{D}} + g^{\ \ \ \; \nu}_{F \dot{D}} g^{ D \dot{D} \mu } ) + \frac{1}{2} (g^{ D \dot{D} \nu} g^{\ \ \ \; \mu}_{ F \dot{D}} - g^{\ \ \ \; \nu}_{F \dot{D}} g^{ D \dot{D} \mu} )
\\ \nonumber
&& = - (g^{\mu \nu} \delta^{D}_{F} + S^{D \ \; \mu \nu}_{\ \; F})
\end{eqnarray}
we obtain
\begin{equation}
\label{pomocnicze_9}
\nabla^{\mu} \nabla_{\mu} C_{ABCF} +  S^{D \ \; \mu \nu}_{\ \; F} \, \nabla_{[\mu} \nabla_{\nu]} C_{ABCD} = \frac{8 \pi G}{c^{4}} \, \nabla_{F}^{\ \; \dot{D}} \nabla_{(A}^{\ \ \; \dot{E}} T_{BC)\dot{D}\dot{E}}
\end{equation}
Using the relation \cite{b12}
\begin{eqnarray}
&& \nabla_{[\mu} \nabla_{\nu]} C_{ABCD} = -2 \, C_{M(ABC} R^{M}_{\ \; D) \mu \nu}, 
\\ \nonumber
&&R^{M}_{\ \; D \mu \nu} = -\frac{1}{2} C^{M}_{\ \; DEF} S^{EF}_{\ \ \ \mu \nu} + \frac{R}{24} \, S^{M}_{\ \; D \mu \nu} + \frac{1}{2} C^{M}_{\ \; D \dot{E} \dot{F}} S^{\dot{E}\dot{F}}_{\ \ \ \mu \nu}
\end{eqnarray}
and performing the straightforward calculations, taking also into account that the right hand side of (\ref{pomocnicze_9}) must be totally symmetric in the indices $(F,A,B,C)$ since the left hand side is totally symmetric in those indices one arrives at the equation
\begin{equation}
\label{pomocnicze_10}
\left( \nabla^{\mu} \nabla_{\mu} + \frac{4 \pi G}{c^{4}} T \right) C_{ABCD} + 6C^{MN}_{\ \ \ (AB} C_{CD)MN} = 
\frac{8 \pi G}{c^{4}} \, \nabla_{(A}^{\ \ \dot{M}} \nabla_{B}^{\ \; \dot{N}} T_{CD) \dot{M}\dot{N}}
\end{equation}
The complex conjugate of Eq. (\ref{pomocnicze_10}) reads
\begin{equation}
\label{pomocnicze_11}
\left( \nabla^{\mu} \nabla_{\mu} + \frac{4 \pi G}{c^{4}} T \right) C_{\dot{A}\dot{B}\dot{C}\dot{D}} + 6C^{\dot{M}\dot{N}}_{\ \ \ (\dot{A}\dot{B}} C_{\dot{C}\dot{D})\dot{M}\dot{N}} = 
\frac{8 \pi G}{c^{4}} \, \nabla^{M}_{\ (\dot{A}} \nabla^{N}_{\ \dot{B}} T_{|MN| \dot{C}\dot{D})}
\end{equation}
Employing (\ref{tensor_matter_spinorial_image}) and the definitions (\ref{nabla_spinorial}) of $\nabla^{A\dot{B}}$, and (\ref{definicja_SAB}) of $S^{AB}_{\ \ \ \mu \nu}$ and $S^{\dot{A} \dot{B}}_{\ \ \ \mu \nu}$ one can rewrite Eqs. (\ref{pomocnicze_10}) and (\ref{pomocnicze_11}) as follows
\begin{equation}
\label{pomocnicze_12}
\left( \nabla^{\mu} \nabla_{\mu} + \frac{4 \pi G}{c^{4}} T \right) C_{ABCD} + 6C^{MN}_{\ \ \ (AB} C_{CD)MN} = \frac{2 \pi G}{c^{4}} \, S_{(AB}^{\ \ \ \ \varrho \mu} S_{CD)}^{\ \ \ \ \sigma \nu} \, \nabla_{\mu} \nabla_{\nu} T_{\varrho \sigma}
\end{equation}
and 
\begin{equation}
\label{pomocnicze_13}
\left( \nabla^{\mu} \nabla_{\mu} + \frac{4 \pi G}{c^{4}} T \right) C_{\dot{A}\dot{B}\dot{C}\dot{D}} + 6C^{\dot{M}\dot{N}}_{\ \ \ (\dot{A}\dot{B}} C_{\dot{C}\dot{D})\dot{M}\dot{N}} = \frac{2 \pi G}{c^{4}} \, S_{(\dot{A}\dot{B}}^{\ \ \ \ \varrho \mu} S_{\dot{C}\dot{D})}^{\ \ \ \ \sigma \nu} \, \nabla_{\mu} \nabla_{\nu} T_{\varrho \sigma}
\end{equation}
Multiplying Eq. (\ref{pomocnicze_12}) by $\dfrac{1}{4} \, S^{AB}_{\ \ \ \alpha \beta} \, S^{CD}_{\ \ \ \gamma \delta}$, Eq. (\ref{pomocnicze_13}) by $\dfrac{1}{4} \, S^{\dot{A}\dot{B}}_{\ \ \ \alpha \beta} \, S^{\dot{C}\dot{D}}_{\ \ \ \gamma \delta}$, then adding together the obtained results and employing (\ref{pomocnicze_15}) we get the equation
\begin{eqnarray}
\label{pomocnicze_16}
&& \left( \nabla^{\mu} \nabla_{\mu} + \frac{4 \pi G}{c^{4}} T \right) C_{\alpha \beta \gamma \delta} + 3 \Re \left\{ S^{AB}_{\ \ \ \alpha \beta} \, S^{CD}_{\ \ \ \gamma \delta} \, C^{MN}_{\ \ \ (AB} C_{CD)MN} \right\} 
\\ \nonumber
&& = \frac{\pi G}{c^{4}} \Re \left\{ S^{AB}_{\ \ \ \alpha \beta} \, S^{CD}_{\ \ \ \gamma \delta} \, S_{(AB}^{\ \ \ \ \varrho \mu} S_{CD)}^{\ \ \ \ \sigma \nu} \right\} \nabla_{\mu} \nabla_{\nu} T_{\varrho \sigma}
\end{eqnarray}
Long and tedious manipulations show that Eq. (\ref{pomocnicze_16}) can be brought to the form
\begin{eqnarray}
\label{pomocnicze_19}
&& \left( \nabla^{\mu} \nabla_{\mu} + \frac{4 \pi G}{c^{4}} T \right) C_{\alpha \beta \gamma \delta} + C_{\alpha \beta \mu \varrho} C^{\mu \varrho}_{\ \ \gamma \delta} + 2 C_{\alpha \mu \varrho \gamma}C^{\ \, \mu \varrho}_{\beta \ \ \; \delta} - 2C_{\alpha \mu \varrho \delta} C^{\ \, \mu \varrho}_{\beta \ \ \; \gamma}
\\ \nonumber
&& = \frac{4\pi G}{c^{4}} \left\{ \frac{1}{3} \left[ (\delta^{\varrho \mu}_{\alpha \beta} \delta^{\sigma \nu}_{\gamma \delta} + \delta^{\sigma \nu}_{\alpha \beta} \delta^{\varrho \mu }_{\gamma \delta})  - |\det (g_{\varkappa \tau})| (\in^{\varrho \mu}_{\bullet \bullet \alpha \beta} \, \in^{\sigma \nu}_{\bullet \bullet \gamma \delta} + \in^{\sigma \nu}_{\bullet \bullet \alpha \beta} \, \in^{\varrho \mu }_{\bullet \bullet \gamma \delta})    \right] \right.
\\ \nonumber
&& \ \ \ \ \ \ \ \  \ \ \ \left. + \frac{1}{6} \, g_{\lambda \iota} g^{\varkappa \tau} \delta^{\eta \iota}_{\varkappa \varepsilon} \, (\delta^{\varrho \mu \lambda}_{\alpha \beta \eta} \delta^{\sigma \nu \varepsilon}_{\gamma \delta \tau} + \delta^{\sigma \nu \varepsilon }_{\alpha \beta \tau} \delta^{\varrho \mu \lambda}_{\gamma \delta \eta}  ) \right\} \nabla_{\mu} \nabla_{\nu} T_{\varrho \sigma}
\end{eqnarray}
where $\in^{\varrho \sigma}_{\bullet \bullet \alpha \beta} = \in_{\mu \nu \alpha \beta}g^{\mu \varrho} g^{\nu \sigma}$, $\delta^{\varrho \mu}_{\alpha \beta}$ and $\delta^{\varrho \mu \lambda}_{\alpha \beta \eta} $ are 2-dimensional and 3-dimensional Cronecker deltas, respectively. Eq. (\ref{pomocnicze_19}) has a rather involved form. One can look for other simpler forms by employing the commutation rule (\ref{komutator_pochodnych_kowariantnych}) but we don't investigate further this problem in the current work. [Another form of (\ref{pomocnicze_19}) is given in \cite{b1} by Eq. (21)].

From Eqs. (\ref{BI_1}) and (\ref{BI_2}) one can easily obtain the following identities
\begin{equation}
\label{pomocnicze_20}
C^{\alpha \beta}_{\ \ \gamma \delta;\alpha \beta}=0
\end{equation}
Eq. (\ref{pomocnicze_20}) is given in a paper by W. Kundt and M. Tr{\"u}mper \cite{b7} where the authors make an important point that if we consider the Weyl tensor as the "free part" of the gravitational field then Eqs. (\ref{matter_equation}) are the equations of motion for the sources, Eqs. (\ref{pomocnicze_2}) describe the interaction between the sources and the free part of the field and Eqs. (\ref{pomocnicze_20}) are differential equations for the free part of the field. We consider this point in the further part of the present work. Analogously, from (\ref{Bianchi_identities_contraction}) one gets \cite{b7}
\begin{equation}
R^{\alpha \beta}_{\ \ \gamma \delta; \alpha \beta} = 0
\end{equation}
To obtain the counterpart of (\ref{pomocnicze_20}) in spinorial formalism we first insert (\ref{pomocnicze_15}) into (\ref{pomocnicze_20}) and then we multiply the formula obtained by $S_{EF}^{\ \ \ \gamma \delta}$ and we employ the relation \cite{b12}
\begin{equation}
S_{EF}^{\ \ \ \gamma \delta} \, S^{\dot{C}\dot{D}}_{\ \ \ \gamma \delta} =0
\end{equation}
Thus one gets
\begin{equation}
\nabla_{\beta} \nabla_{\alpha} ( S^{AB \alpha \beta} \, S^{CD}_{\ \ \ \gamma \delta} \, S_{EF}^{\ \ \ \gamma \delta} \, C_{ABCD} ) = 0
\end{equation}
Using the identity \cite{b12}
\begin{equation}
S^{CD}_{\ \ \ \gamma \delta} \, S_{EF}^{\ \ \ \gamma \delta} = 8 \, \delta^{C}_{(E} \, \delta^{D}_{F)}
\end{equation}
we have 
\begin{equation}
\nabla_{\beta} \nabla_{\alpha} (S^{AB \alpha \beta} C_{ABEF} )=0
\end{equation}
Finally, employing the definition (\ref{definicja_SAB}) of $S^{AB}$ and the definition (\ref{nabla_spinorial}) of $\nabla^{A \dot{B}}$ one arrives at the equation
\begin{equation}
\label{pomocnicze_21}
\nabla^{A}_{\ \dot{M}} \nabla^{B \dot{M}} C_{ABEF}=0
\end{equation}
It can be easily shown that (\ref{pomocnicze_21}) is equivalent to (\ref{pomocnicze_20}). The complex conjugate of (\ref{pomocnicze_21}) reads
\begin{equation}
\nabla_{M}^{\ \; \dot{A}} \nabla^{M \dot{B}} C_{\dot{A}\dot{B}\dot{E}\dot{F}}=0
\end{equation}


\section{Gravito-electromagnetism in terms of Pleba{\'n}ski's helicity formalism}
\setcounter{equation}{0}

A close analogy between electromagnetism and gravity has been investigated by many authors (see for example \cite{b14, b15, b16, b17, b18, b19, b20, b21, b22}). In 1980 J.F. Pleba{\'n}ski \cite{b3} developed the \textsl{helicity formalism} for Riemannian structures in complex or real four dimensions. This formalism provides us with a powerful and elegant tool for presentation the so called gravito-electromagnetism \cite{b21} i.e. the correspondence between electromagnetism and Einstein gravity.

Define three $2$x$2$ matrices $(\Phi_{a}^{\ AB})$, $a=1,2,3$
\begin{equation}
\label{definicja_macierzy_Phi}
(\Phi_{a}^{\ AB}) = \left\{
 \frac{1}{\sqrt{2}} \left(\begin{array}{cc}
i & \ \ 0 \\
0 & -i
\end{array}\right), \ 
 \frac{1}{\sqrt{2}} \left(\begin{array}{cc}
1 & \ \ 0 \\
0 & \ \ 1
\end{array}\right), \ 
 \frac{1}{\sqrt{2}} \left(\begin{array}{cc}
\ \ 0 & -i \\
-i & \ \ 0
\end{array}\right) \right\}
\end{equation}
and their complex conjugate $(\Phi_{\dot{a}}^{\ \dot{A}\dot{B}})$, $\dot{a} = \dot{1}, \dot{2}, \dot{3}$
\begin{equation}
\label{definicja_macierzy_Phi_complex_conjugate}
(\Phi_{\dot{a}}^{\ \dot{A}\dot{B}}) = \left\{
 \frac{1}{\sqrt{2}} \left(\begin{array}{cc}
-i & \ 0 \\
\ \ 0 & \ i
\end{array}\right), \ 
 \frac{1}{\sqrt{2}} \left(\begin{array}{cc}
1 & \ \ 0 \\
0 & \ \ 1
\end{array}\right), \ 
 \frac{1}{\sqrt{2}} \left(\begin{array}{cc}
0 & \ i \\
i & \ 0
\end{array}\right) \right\}
\end{equation}
(Note that in \cite{b3} the respective matrices $(\Phi_{a}^{\ AB})$ and $(\Phi_{\dot{a}}^{\ \dot{A}\dot{B}})$ do not have the factor $\dfrac{1}{\sqrt{2}}$). 

One quickly gets the relations
\begin{eqnarray}
\label{zaleznosci_helicity}
&& \Phi_{a}^{\ AB} \, \Phi_{bAB} = \delta_{ab}, \ \ \Phi_{a}^{\ AB} \, \Phi^{a}_{\ CD} = \delta^{A}_{(C} \delta^{B}_{D)}
\\ \nonumber
&& \Phi_{\dot{a}}^{\ \dot{A}\dot{B}} \, \Phi_{\dot{b} \dot{A} \dot{B}} = \delta_{\dot{a}\dot{b}}, \ \ \Phi_{\dot{a}}^{\ \dot{A}\dot{B}} \, \Phi^{\dot{a}}_{\ \dot{C}\dot{D}} = \delta^{\dot{A}}_{(\dot{C}} \delta^{\dot{B}}_{\dot{D})}
\end{eqnarray}
where the spinorial indices $A,B,...,\dot{A}, \dot{B},...$ are to be manipulated according to (\ref{spinorowe_podnoszenie_wskaznikow}) and the "helicity indices" $a,b,\dot{a}, \dot{b}$ are to be manipulated with the use of the Kronecker delta.

Let $\Psi^{AB} = \Psi^{(AB)}$ be a symmetric spinor of the second rank at some point $p$ of the spacetime $\mathcal{M}$. One can assign to $\Psi^{AB}$ a complex vector $(\Psi^{1}, \Psi^{2}, \Psi^{3}) \in \mathbb{C}^{3}$ by the formula
\begin{equation}
\label{skladowe_wektora_Psi}
\Psi^{a} = \Phi^{a}_{\ AB} \, \Psi^{AB}, \ a=1,2,3
\end{equation}
In a new spinor basis
\begin{equation}
\nonumber
e'_{A} = l^{-1 \, B}_{\ \ \ \  A} \, e_{B}
\end{equation}
the components of the spinor $\Psi^{AB}$ are
\begin{equation}
\label{transformacja_Psi}
\Psi'^{AB} = l^{A}_{\ C} \, l^{B}_{\ D} \, \Psi^{CD}, \ \ (l^{A}_{\ C}) \in SL(2; \mathbb{C}), \ \ l^{A}_{\ C} \, l^{-1 \, C}_{\ \ \ \ D} = \delta^{A}_{D}
\end{equation}
Then according to (\ref{skladowe_wektora_Psi}) we define $(\Psi'^{1}, \Psi'^{2}, \Psi'^{3}) \in \mathbb{C}^{3}$ as
\begin{equation}
\label{definicja_Psi_prim}
\Psi'^{a} = \Phi^{a}_{\ AB} \, \Psi'^{AB}
\end{equation}
Inserting (\ref{transformacja_Psi}) into (\ref{definicja_Psi_prim}) and employing the relation which follows from (\ref{skladowe_wektora_Psi}) and (\ref{zaleznosci_helicity})
\begin{equation}
\Psi^{AB} = \Phi_{a}^{\ AB} \, \Psi^{a}
\end{equation}
one obtains
\begin{eqnarray}
\label{pomocnicze_23}
&& \Psi'^{a} = \Phi^{a}_{\ AB} \, l^{A}_{\ C} \, l^{B}_{\ D} \, \Psi^{CD} = \Phi^{a}_{\ AB} \, l^{A}_{\ C} \, l^{B}_{\ D} \, \Phi_{b}^{\ CD} \, \Psi^{b} = t^{a}_{\ b} \, \Psi^{b}
\\ \nonumber
&& t^{a}_{\ b} = \Phi^{a}_{\ AB} \, l^{A}_{\ C} \, l^{B}_{\ D} \, \Phi_{b}^{\ CD} 
\end{eqnarray}
We quickly find that $(t^{a}_{\ b}) \in SO(3; \mathbb{C})$. Moreover, given $(t^{a}_{\ b}) \in SO(3; \mathbb{C})$ the matrix $(l^{A}_{\ B}) \in SL(2; \mathbb{C})$ is defined uniquely up to the sign. Therefore the relation between $(l^{A}_{\ B})$ and $(t^{a}_{\ b})$ given by (\ref{pomocnicze_23}) defines the group isomorphism 
\begin{equation}
\label{izomorfizm}
\quotientfour{SL(2; \mathbb{C})}{\mathbb{Z}_{2}} = SO(3; \mathbb{C})
\end{equation}
Writing the last formula of (\ref{pomocnicze_23}) in the form
\begin{equation}
t^{a}_{\ b} = \Phi^{a}_{\ AB} \, l^{(A}_{\ (C} \, l^{B)}_{\ D)} \, \Phi_{b}^{\ CD}
\end{equation}
we get the group isomorphism
\begin{equation}
SL(2; \mathbb{C}) \underset{s}{\otimes} SL (2 ; \mathbb{C})  = SO(3; \mathbb{C})
\end{equation}
Finally, since the group $\quotientfour{SL(2; \mathbb{C})}{\mathbb{Z}_{2}}$ is isomorphic to the proper ortochronous  Lorentz group $SO^{\uparrow} (1,3)$ then by (\ref{izomorfizm}) one has
\begin{equation}
SO^{\uparrow} (1,3) = SO(3; \mathbb{C})
\end{equation}
The analogous construction can be done for the dotted symmetric spinor $\Psi^{\dot{A}\dot{B}} = \Psi^{(\dot{A}\dot{B})} = \overline{\Psi^{AB}}$. We define
\begin{equation}
\label{definicja_dotted_helicity}
\Psi^{\dot{a}}= \Phi^{\dot{a}}_{\ \dot{A}\dot{B}} \, \Psi^{\dot{A}\dot{B}} = \overline{\Psi^{a}}, \ \ \ \ \dot{a} = \dot{1}, \dot{2}, \dot{3}
\end{equation}
In a new spinor basis
\begin{equation}
e'_{\dot{A}} = l^{-1 \, \dot{B}}_{\ \ \ \ \dot{A}} \, e_{\dot{B}}, \ \ \ \  l^{-1 \, \dot{B}}_{\ \ \ \ \dot{A}} = \overline{l^{-1 \, B}_{\ \ \ \ A}}
\end{equation}
we have
\begin{equation}
\Psi'^{\dot{A}\dot{B}} = l^{\dot{A}}_{\ \dot{C}} \, l^{\dot{B}}_{\ \dot{D}} \, \Psi^{\dot{C}\dot{D}}, \ \ \ \overline{(l^{A}_{\ C})} = (l^{\dot{A}}_{\ \dot{C}}) \in SL(2; \mathbb{C})
\end{equation}
Then
\begin{eqnarray}
&& \Psi'^{\dot{a}} = \Phi^{\dot{a}}_{\ \dot{A}\dot{B}} \Psi'^{\dot{A}\dot{B}} = \Phi^{\dot{a}}_{\ \dot{A}\dot{B}} \, l^{\dot{A}}_{\ \dot{C}} \, l^{\dot{B}}_{\ \dot{D}} \, \Phi_{\dot{b}}^{\ \dot{C} \dot{D}} \, \Psi^{\dot{b}} = t^{\dot{a}}_{\ \dot{b}} \, \Psi^{\dot{b}}
\\ \nonumber
&& t^{\dot{a}}_{\ \dot{b}} = \Phi^{\dot{a}}_{\ \dot{A}\dot{B}} \, l^{\dot{A}}_{\ \dot{C}} \, l^{\dot{B}}_{\ \dot{D}} \, \Phi_{\dot{b}}^{\ \dot{C}\dot{D}} = \overline{t^{a}_{\ b}}
\end{eqnarray}
Of course, $(t^{\dot{a}}_{\ \dot{b}}) \in \overline{SO} (3; \mathbb{C})$. 

It is quite clear that one can generalize the procedure describing above on the case of any spinor $\Psi^{A_{1}B_{1}...A_{n}B_{n}\dot{A}_{1}\dot{B}_{1}...\dot{A}_{m}\dot{B}_{m}}$ symmetric for the indices $(A_{1}B_{1}),...,(A_{n}B_{n})$ and $(\dot{A}_{1}\dot{B}_{1}),...,(\dot{A}_{m}\dot{B}_{m})$
\begin{eqnarray}
\label{wielokrotne_spinory}
&& \Psi^{A_{1}B_{1}...A_{n}B_{n}\dot{A}_{1}\dot{B}_{1}...\dot{A}_{m}\dot{B}_{m}} = 
   \Psi^{(A_{1}B_{1})...A_{n}B_{n}\dot{A}_{1}\dot{B}_{1}...\dot{A}_{m}\dot{B}_{m}} = 
 \\ \nonumber
 && \Psi^{A_{1}B_{1}...(A_{n}B_{n})\dot{A}_{1}\dot{B}_{1}...\dot{A}_{m}\dot{B}_{m}} = 
   \Psi^{A_{1}B_{1}...A_{n}B_{n}(\dot{A}_{1}\dot{B}_{1})...\dot{A}_{m}\dot{B}_{m}} = 
   \\ \nonumber
  && \Psi^{A_{1}B_{1}...A_{n}B_{n}\dot{A}_{1}\dot{B}_{1}...(\dot{A}_{m}\dot{B}_{m})} 
\end{eqnarray}
by assigning to such a spinor the following tensor
\begin{equation}
\Psi^{a_{1}...a_{n}\dot{a}_{1}...\dot{a}_{m}} = \Phi^{a_{1}}_{\ A_{1}B_{1}}...\Phi^{a_{n}}_{\ A_{n} B_{n}} \Phi^{\dot{a}_{1}}_{\ \dot{A}_{1} \dot{B}_{1}}...\Phi^{\dot{a}_{m}}_{\ \dot{A}_{m}\dot{B}_{m}} \, \Psi^{A_{1}B_{1}...A_{n}B_{n}\dot{A}_{1}\dot{B}_{1}...\dot{A}_{m}\dot{B}_{m}}
\end{equation}
The rule of transformation of this tensor reads
\begin{equation}
\Psi'^{a_{1}...a_{n}\dot{a}_{1}...\dot{a}_{m}} = t^{a_{1}}_{\ b_{1}}...t^{a_{n}}_{\ b_{n}}t^{\dot{a}_{1}}_{\ \dot{b}_{1}}...t^{\dot{a}_{m}}_{\ \dot{b}_{m}} \, \Psi^{b_{1}...b_{n}\dot{b}_{1}...\dot{b}_{m}}
\end{equation}
The analogous correspondence can be done for mixed spinors. For example for $\Psi^{A_{1}B_{1}}_{\ \ \ \ \dot{A}_{1}\dot{B}_{1}}$ we define
\begin{equation}
\Psi^{a_{1}}_{\ \dot{a}_{1}} = \Phi^{a_{1}}_{\ A_{1}B_{1}} \Phi_{\dot{a}_{1}}^{\ \; \dot{A}_{1}\dot{B}_{1}} \, \Psi^{A_{1}B_{1}}_{\ \ \ \ \dot{A}_{1}\dot{B}_{1}}
\end{equation}
with the transformation rule
\begin{equation}
\label{primowane_transformacja_pomocnicze}
\Psi'^{a_{1}}_{\ \ \dot{a}_{1}} = t^{a_{1}}_{\ b_{1}} t^{-1 \, \dot{b}_{1}}_{\ \ \ \ \dot{a}_{1}} \, \Psi^{b_{1}}_{\ \dot{b}_{1}}
\end{equation}
From the geometrical point of view one has a clear picture. The formula (\ref{skladowe_wektora_Psi}) defines an isomorphism between the vector bundle of the symmetric contravariant undotted spinors of the second rank over the spacetime $\mathcal{M}$ and the vector bundle of the standard fibre $\mathbb{C}^{3}$and the structure group $SO(3;\mathbb{C})$ also over $\mathcal{M}$. Similarly the formula (\ref{definicja_dotted_helicity}) defines an isomorphism between the vector bundle of the symmetric contravariant dotted spinors of the second rank over $\mathcal{M}$ and the vector bundle of the standard fibre $\mathbb{C}^{3}$ and the structure group $\overline{SO}(3; \mathbb{C})$. Further generalizations described by the formulae (\ref{wielokrotne_spinory}) to (\ref{primowane_transformacja_pomocnicze}) give the isomorphism between the tensor products of the spinorial vector bundles and the respective tensor products of the vector bundles with the standard fibre $\mathbb{C}^{3}$ and the structure group $SO(3;\mathbb{C})$ or $\overline{SO}(3;\mathbb{C})$. This observation enables us to translate the spinor formalism into the helicity formalism. Consequently, the matrices $(\Phi_{a}^{\ AB})$ and $(\Phi_{\dot{a}}^{\ \dot{A}\dot{B}})$ are the \textsl{soldering objects} which connect spinors with respective $SO(3;\mathbb{C})$ (and $\overline{SO}(3;\mathbb{C})$) tensors analogously as the matrices $(g_{\ \ \ \mu}^{A \dot{B}})$ given by (\ref{definicja_g_mu_AB}) connect the spacetime tensors with corresponding spinors. Then the \textsl{helicity formalism} arises as the formalism on the $SO(3;\mathbb{C})$ (and $\overline{SO}(3;\mathbb{C})$) tensor bundles by the isomorphism described above. 

We have the following basic correspondences
\begin{subequations}
\begin{eqnarray}
\label{korespondencja_1}
&& S^{AB} \ \longleftrightarrow \ \mathcal{S}^{a} = \Phi^{a}_{\ AB} \, S^{AB}, \ S^{\dot{A}\dot{B}} \ \longleftrightarrow \ \mathcal{S}^{\dot{a}} = \Phi^{\dot{a}}_{\ \dot{A}\dot{B}} \, S^{\dot{A}\dot{B}} = \overline{\mathcal{S}^{a}}
\\ 
\label{korespondencja_2}
&& C_{ABCD} \ \longleftrightarrow \ \mathcal{C}_{ab} = \Phi_{a}^{\ AB} \, \Phi_{b}^{\ CD} \, C_{ABCD} = \mathcal{C}_{(ab)}
\\ \nonumber
&& C_{\dot{A}\dot{B}\dot{C}\dot{D}} \ \longleftrightarrow \ \mathcal{C}_{\dot{a}\dot{b}} = \Phi_{\dot{a}}^{\ \dot{A}\dot{B}} \, \Phi_{\dot{b}}^{\ \dot{C}\dot{D}} \, C_{\dot{A}\dot{B}\dot{C}\dot{D}} = \mathcal{C}_{(\dot{a}\dot{b})} = \overline{\mathcal{C}_{ab}}
\\
\label{korespondencja_3}
&& C_{AB \dot{C}\dot{D}} \ \longleftrightarrow \ \mathcal{C}_{a\dot{b}} = \Phi_{a}^{\ AB} \, \Phi_{\dot{b}}^{\ \dot{C}\dot{D}} \, C_{AB \dot{C}\dot{D}} = \overline{\mathcal{C}_{b \dot{a}}}
\\
&& R \ \longleftrightarrow \ R
\end{eqnarray}
\end{subequations}
The \textsl{exterior covariant differentiation $D$} in the helicity formalism is given by
\begin{eqnarray}
\label{exterior_derivative_definition}
\mathcal{D} \Psi^{a_{1}...a_{n}\dot{a}_{1}...\dot{a}_{m}}_{b_{1}...b_{r}\dot{b}_{1}...\dot{b}_{s}} &=&
\Phi^{a_{1}}_{\ \ A_{1}B_{1}}...\Phi^{a_{n}}_{\ \ A_{n}B_{n}} \, \Phi^{\dot{a}_{1}}_{\ \ \dot{A}_{1}\dot{B}_{1}}...\Phi^{\dot{a}_{m}}_{\ \ \dot{A}_{m}\dot{B}_{m}} \, \Phi_{b_{1}}^{\ \ C_{1}D_{1}}...\Phi_{b_{r}}^{\ \ C_{r}D_{r}} 
\\ \nonumber
&&  \Phi_{\dot{b}_{1}}^{\ \ \dot{C}_{1}\dot{D}_{1}}...\Phi_{\dot{b}_{s}}^{\ \ \dot{C}_{s}\dot{D}_{s}} \, D\Psi^{A_{1}B_{1}...A_{n}B_{n}\dot{A}_{1}\dot{B}_{1}...\dot{A}_{m}\dot{B}_{m}}_{C_{1}D_{1}...C_{r}D_{r}\dot{C}_{1}\dot{D}_{1}...\dot{C}_{s}\dot{D}_{s}}
\\ \nonumber
&=& d \Psi^{a_{1}...a_{n}\dot{a}_{1}...\dot{a}_{m}}_{b_{1}...b_{r}\dot{b}_{1}...\dot{b}_{s}}  + \Gamma^{a_{1}}_{\ c_{1}} \Psi^{c_{1}...a_{n}\dot{a}_{1}...\dot{a}_{m}}_{b_{1}...b_{r}\dot{b}_{1}...\dot{b}_{s}} +...+ \Gamma^{a_{n}}_{\ c_{n}} \Psi^{a_{1}...c_{n}\dot{a}_{1}...\dot{a}_{m}}_{b_{1}...b_{r}\dot{b}_{1}...\dot{b}_{s}} 
\\ \nonumber
&&  \ \ \ \ \ \ \ \ \ \ \ \ \ \ \ \ \ \
+ \ \Gamma^{\dot{a}_{1}}_{\ \dot{c}_{1}} \Psi^{a_{1}...a_{n}\dot{c}_{1}...\dot{a}_{m}}_{b_{1}...b_{r}\dot{b}_{1}...\dot{b}_{s}} +...+ \Gamma^{\dot{a}_{m}}_{\ \dot{c}_{m}} \Psi^{a_{1}...a_{n}\dot{a}_{1}...\dot{c}_{m}}_{b_{1}...b_{r}\dot{b}_{1}...\dot{b}_{s}} 
\\ \nonumber
&& \ \ \ \ \ \ \ \ \ \ \ \ \ \ \ \ \ \
- \ \Gamma^{c_{1}}_{\ b_{1}} \Psi^{a_{1}...a_{n}\dot{a}_{1}...\dot{a}_{m}}_{c_{1}...b_{r}\dot{b}_{1}...\dot{b}_{s}}+...
-\Gamma^{c_{r}}_{\ b_{r}} \Psi^{a_{1}...a_{n}\dot{a}_{1}...\dot{a}_{m}}_{b_{1}...c_{r}\dot{b}_{1}...\dot{b}_{s}}
\\ \nonumber
&&  \ \ \ \ \ \ \ \ \ \ \ \ \ \ \ \ \ \
- \ \Gamma^{\dot{c}_{1}}_{\ \dot{b}_{1}}  \Psi^{a_{1}...a_{n}\dot{a}_{1}...\dot{a}_{m}}_{b_{1}...b_{r}\dot{c}_{1}...\dot{b}_{s}} +...
-\Gamma^{\dot{c}_{s}}_{\ \dot{b}_{s}}  \Psi^{a_{1}...a_{n}\dot{a}_{1}...\dot{a}_{m}}_{b_{1}...b_{r}\dot{b}_{1}...\dot{c}_{s}}
\end{eqnarray}
where $\Gamma^{a}_{\ b}$ and $\Gamma^{\dot{a}}_{\ \dot{b}}$ are related to the spinorial connection forms $\Gamma^{A}_{\ B}$ and $\Gamma^{\dot{A}}_{\ \dot{B}}$ as follows
\begin{equation}
\label{zwiazek_miedzy_koneksjami}
\Gamma^{a}_{\ b} = 2 \Phi^{a}_{\ AC} \, \Phi_{b}^{\ CB} \, \Gamma^{A}_{\ B}, \ \Gamma^{\dot{a}}_{\ \dot{b}} = 2 \Phi^{\dot{a}}_{\ \dot{A}\dot{C}} \, \Phi_{\dot{b}}^{\ \dot{C} \dot{B}}\, \Gamma^{\dot{A}}_{\ \dot{B}} = \overline{\Gamma^{a}_{\ b}}
\end{equation}
One can easily check that 
\begin{equation}
\label{pomocnicze_24}
\Phi_{aDA} \, \Phi^{\  A}_{b \ \; B} \, \Phi_{c}^{\ BD} = \frac{1}{\sqrt{2}} \in_{abc}
\end{equation}
Using (\ref{pomocnicze_24}) and employing also (\ref{zaleznosci_helicity}) we rewrite (\ref{zwiazek_miedzy_koneksjami}) as
\begin{equation}
\label{pomocnicze_26}
\Gamma^{a}_{\ b} = \sqrt{2} \in_{acb} \Gamma^{c}, \ \Gamma^{\dot{a}}_{\ \dot{b}} = \sqrt{2} \in_{\dot{a}\dot{c}\dot{b}} \Gamma^{\dot{c}}, \ \Gamma^{c} = \Phi^{cAB} \, \Gamma_{AB} = \overline{\Gamma^{\dot{c}}}
\end{equation}
One can easily show that in the orthonormal basis in which (\ref{definicja_g_mu_AB}) holds true we have the following important relation between $S^{AB}$ and $\Phi_{a}^{\ AB}$ ($S^{\dot{A}\dot{B}}_{\ \ \ \; 0a}$ and $\Phi_{\dot{a}}^{\ \dot{A}\dot{B}}$), $a=1,2,3$, $\dot{a} = \dot{1}, \dot{2}, \dot{3}$
\begin{equation}
\label{pomocnicze_25}
S^{AB}_{\ \ \ \; 0a} = i \sqrt{2} \, \Phi_{a}^{\ AB}, \ S^{\dot{A}\dot{B}}_{\ \ \ \; 0a} = - i \sqrt{2} \, \Phi_{\dot{a}}^{\ \dot{A}\dot{B}}
\end{equation}
where $S^{AB}$ and $S^{\dot{A}\dot{B}}$ are given by (\ref{definicja_SAB}). Employing (\ref{pomocnicze_25}), (\ref{pomocnicze_15}) and (\ref{pomocnicze_15_a}) in (\ref{korespondencja_2}) we obtain
\begin{eqnarray}
\label{wzory_na_Cab}
\mathcal{C}_{ab} &=& -\frac{1}{2} S^{AB}_{\ \ \ \; 0a} S^{CD}_{\ \ \ \; 0b} \, C_{ABCD} = -(C_{0a0b} + \ast C_{0a0b})
\\ \nonumber
\mathcal{C}_{\dot{a}\dot{b}} &=& -\frac{1}{2} S^{\dot{A}\dot{B}}_{\ \ \ \; 0a}S^{\dot{C}\dot{D}}_{\ \ \ \; 0b} \, C_{\dot{A}\dot{B}\dot{C}\dot{D}} = - (C_{0a0b} - \ast C_{0a0b})
\end{eqnarray}
Define the \textsl{electric part $\mathcal{E}_{ab}$} and \textsl{magnetic part $\mathcal{B}_{ab}$} of the free gravitational field represented by the Weyl tensor $C_{\alpha \beta \gamma \delta}$
\begin{equation}
\label{electric_magnetic_curvature}
\mathcal{E}_{ab} = C_{0a0b} = \mathcal{E}_{(ab)}, \ \mathcal{B}_{ab} = i \ast C_{0a0b} = \mathcal{B}_{(ab)}, \ a,b=1,2,3
\end{equation}
(compare with \cite{b1, b2, b5, b18, b21, b23}). Then (\ref{wzory_na_Cab}) can be rewritten as 
\begin{equation}
\label{zaleznosc_miedzy_C_E_B}
\mathcal{C}_{ab} = - \mathcal{E}_{ab} + i \mathcal{B}_{ab}, \ \mathcal{C}_{\dot{a}\dot{b}} = -(\mathcal{E}_{ab} + i \mathcal{B}_{ab})
\end{equation}
The 3x3 matrices $(\mathcal{E}_{ab})$ and $(\mathcal{B}_{ab})$ are real, symmetric and traceless.

From the spinorial relations \cite{b3, b12}
\begin{eqnarray}
DS^{AB} &=& d S^{AB} + \Gamma^{A}_{\ C} \wedge S^{CB} + \Gamma^{B}_{\ C} \wedge S^{AC}=0
\\ \nonumber
DS^{\dot{A}\dot{B}} &=& dS^{\dot{A}\dot{B}} + \Gamma^{\dot{A}}_{\ \dot{C}} \wedge S^{\dot{C}\dot{B}} + \Gamma^{\dot{B}}_{\ \dot{C}} \wedge S^{\dot{A}\dot{C}}=0
\end{eqnarray}
by (\ref{korespondencja_1}), (\ref{exterior_derivative_definition}) and (\ref{pomocnicze_26}) we obtain
\begin{eqnarray}
\mathcal{D} \mathcal{S}^{a} &=& d \mathcal{S}^{a} + \Gamma^{a}_{\ c} \wedge \mathcal{S}^{c} = d \mathcal{S}^{a} + \sqrt{2} \in_{abc} \Gamma^{b} \wedge \mathcal{S}^{c}=0
\\ \nonumber
\mathcal{D} \mathcal{S}^{\dot{a}} &=& d \mathcal{S}^{\dot{a}} + \Gamma^{\dot{a}}_{\ \dot{c}} \wedge \mathcal{S}^{\dot{c}} = d \mathcal{S}^{\dot{a}} + \sqrt{2} \in_{\dot{a}\dot{b}\dot{c}} \Gamma^{\dot{b}} \wedge \mathcal{S}^{\dot{c}}=0
\end{eqnarray}
The second Cartan structure equations in spinorial language read
\begin{eqnarray}
R^{A}_{\ B} &=& d \Gamma^{A}_{\ B} + \Gamma^{A}_{\ C} \wedge \Gamma^{C}_{\ B} = -\frac{1}{2} \, C^{A}_{\ BCD} S^{CD} + \frac{R}{24} \, S^{A}_{\ B} + \frac{1}{2} \, C^{A}_{\ B \dot{C}\dot{D}} S^{\dot{C}\dot{D}}
\\ \nonumber
R^{\dot{A}}_{\ \dot{B}} &=& d \Gamma^{\dot{A}}_{\ \dot{B}} + \Gamma^{\dot{A}}_{\ \dot{C}} \wedge \Gamma^{\dot{C}}_{\ \dot{B}} = - \frac{1}{2} \, C^{\dot{A}}_{\ \dot{B}\dot{C}\dot{D}} S^{\dot{C}\dot{D}} + \frac{R}{24} \, S^{\dot{A}}_{\ \dot{B}} + \frac{1}{2} \, C_{CD \ \dot{B}}^{\ \ \ \, \dot{A}} S^{CD}
\end{eqnarray}
where $R^{A}_{\ B} = \dfrac{1}{2} \, R^{A}_{\ B \mu \nu} \, dx^{\mu} \wedge dx^{\nu}$ and $R^{\dot{A}}_{\ \dot{B}} := \dfrac{1}{2} \, R^{\dot{A}}_{\ \dot{B} \mu \nu} \, dx^{\mu} \wedge dx^{\nu} = \overline{R^{A}_{\ B}}$ are the spinorial curvature 2-forms. In helicity formalism one quickly gets the second Cartan structure equations in the form
\begin{eqnarray}
\label{curvature_forms_helicity}
\mathcal{R}^{a} &=& d \Gamma^{a} + \frac{1}{\sqrt{2}} \in_{abc} \Gamma^{b} \wedge \Gamma^{c} = -\frac{1}{2} \, \mathcal{C}^{a}_{\ b} \mathcal{S}^{b} + \frac{R}{24} \, \mathcal{S}^{a} + \frac{1}{2} \, \mathcal{C}^{a}_{\ \dot{b}} \mathcal{S}^{\dot{b}}
\\ \nonumber
\mathcal{R}^{\dot{a}} &=& d \Gamma^{\dot{a}} + \frac{1}{\sqrt{2}} \in_{\dot{a}\dot{b}\dot{c}} \Gamma^{\dot{b}} \wedge \Gamma^{\dot{c}} = -\frac{1}{2} \, \mathcal{C}^{\dot{a}}_{\ \dot{b}} \mathcal{S}^{\dot{b}} + \frac{R}{24} \, \mathcal{S}^{\dot{a}} + \frac{1}{2} \, \mathcal{C}_{b}^{\ \dot{a}} \mathcal{S}^{b} = \overline{\mathcal{R}^{a}}
\end{eqnarray}
where $\mathcal{R}^{a} = \Phi^{aAB}R_{AB}$ and $\mathcal{R}^{\dot{a}} = \Phi^{\dot{a}\dot{A}\dot{B}} R_{\dot{A}\dot{B}} = \overline{\mathcal{R}^{a}}$.

Now the crucial point is to study the Bianchi identities since in the present approach to the curvature dynamics the Bianchi identities are considered as the field equations. The Bianchi identities in helicity formalism read
\begin{eqnarray}
\label{Bianchi_identities_helicity}
\mathcal{D} \mathcal{R}^{a} &=& d \mathcal{R}^{a} + \sqrt{2} \in_{abc} \Gamma^{b} \wedge \mathcal{R}^{c} = 0
\\ \nonumber
\mathcal{D} \mathcal{R}^{\dot{a}} &=& d \mathcal{R}^{\dot{a}} + \sqrt{2} \in_{\dot{a}\dot{b}\dot{c}} \Gamma^{\dot{b}} \wedge \mathcal{R}^{\dot{c}} = \overline{ \mathcal{D}\mathcal{R}^{a}} = 0
\end{eqnarray}
Inserting (\ref{curvature_forms_helicity}) into (\ref{Bianchi_identities_helicity}) we get
\begin{eqnarray}
\label{pomocnicze_27}
&& - \mathcal{D} \mathcal{C}^{a}_{\ b} \wedge \mathcal{S}^{b} + \frac{1}{12} \, dR \wedge \mathcal{S}^{a} + \mathcal{D} \mathcal{C}^{a}_{\ \dot{b}} \wedge \mathcal{S}^{\dot{b}} = 0
\\ \nonumber
&& -\mathcal{D} \mathcal{C}^{\dot{a}}_{\ \dot{b}} \wedge \mathcal{S}^{\dot{b}} + \frac{1}{12} \, dR \wedge \mathcal{S}^{\dot{a}} + \mathcal{D} \mathcal{C}_{b}^{\ \dot{a}} \wedge \mathcal{S}^{b} =0
\end{eqnarray}
From (\ref{pomocnicze_6}) and (\ref{korespondencja_3}) one finds
\begin{equation}
\mathcal{C}_{a \dot{b}} = - \frac{8\pi G}{c^{4}} \, \mathcal{T}_{a \dot{b}}, \ \mathcal{T}_{a\dot{b}} := \Phi_{a}^{\ AB} \, \Phi_{\dot{b}}^{\ \dot{C}\dot{D}} \, T_{AB \dot{C}\dot{D}}
\end{equation}
Employing also (\ref{matter_skalar}) we conclude that the identities (\ref{pomocnicze_27}) lead to the following equations
\begin{eqnarray}
\label{Bianchi_dentities_another_form}
\mathcal{D} \mathcal{C}^{a}_{\ b} \wedge \mathcal{S}^{b} &=& \frac{8 \pi G}{c^{4}} \left( - \mathcal{D}\mathcal{T}^{a}_{\ \; \dot{b}} \wedge \mathcal{S}^{\dot{b}} + \frac{1}{12} \, dT \wedge \mathcal{S}^{a} \right)
\\ \nonumber
\mathcal{D} \mathcal{C}^{\dot{a}}_{\ \dot{b}} \wedge \mathcal{S}^{\dot{b}} &=& \frac{8 \pi G}{c^{4}} \left( - \mathcal{D} \mathcal{T}_{b}^{\ \dot{a}} \wedge \mathcal{S}^{b} + \frac{1}{12} \, dT \wedge \mathcal{S}^{\dot{a}} \right)
\end{eqnarray}
Eqs. (\ref{Bianchi_dentities_another_form}) are the field equations for the curvature dynamics in the helicity formalism.

Note that by (\ref{exterior_derivative_definition}) and (\ref{pomocnicze_26})
\begin{equation}
\mathcal{D} \mathcal{C}^{a}_{\ b} = d \mathcal{C}^{a}_{\ b} + \sqrt{2} \in_{acd} \Gamma^{c} \mathcal{C}^{d}_{\ b} + \sqrt{2} \in_{bcd} \Gamma^{c} \mathcal{C}^{ad}
\end{equation}
and analogously for other objects of (\ref{Bianchi_dentities_another_form}). Eqs. (\ref{Bianchi_dentities_another_form}) are the main field equations of curvature dynamics in helicity formalism. By the relations (\ref{zaleznosc_miedzy_C_E_B}) Eqs. (\ref{Bianchi_dentities_another_form}) can be rewritten in terms of the electric part $\mathcal{E}_{ab}$ and the magnetic part $\mathcal{B}_{ab}$ of the gravitational field. Thus our equations (\ref{Bianchi_dentities_another_form}) are the helicity counterparts of the respective equations given in \cite{b18, b21}. We show that Eqs. (\ref{Bianchi_dentities_another_form}) are the gravitational analogous of the electromagnetic field equations. Maxwell equations in the presence of currents $j^{\mu}$ are \cite{b12}
\begin{equation}
\label{Maxwell_equations}
f^{\mu\nu}_{\ \ ; \nu} = \frac{4 \pi}{c} \, j^{\mu} , \ \ast f^{\mu\nu}_{\ \ ; \nu} = 0
\end{equation}
where $f_{\mu \nu}=-f_{\nu \mu}$ is the tensor of electromagnetic fields
\begin{equation}
f_{\mu \nu} = \frac{\partial A_{\nu}}{\partial x^{\mu}} - \frac{\partial A_{\mu}}{\partial x^{\nu}}
\end{equation}
with $A_{\mu}$ denoting the electromagnetic potential vector and 
\begin{equation}
\ast f^{\mu \nu} = -\frac{1}{2} i | \det(g_{\alpha \beta})|^{-\frac{1}{2}} \in^{\mu \nu \gamma \delta} f_{\gamma \delta}
\end{equation}
Simple manipulations show that the first part of the Maxwell equations (\ref{Maxwell_equations}) can be equivalently written as
\begin{equation}
(\ast f_{[\alpha \beta} )_{; \gamma]} = \frac{4 \pi}{3c} i | \det(g_{\mu \nu})|^{\frac{1}{2}} \in_{\delta \alpha \beta \gamma}  j^{\delta}
\end{equation}
or inserting
\begin{equation}
\ast j_{\alpha \beta \gamma} = | \det(g_{\mu \nu})|^{\frac{1}{2}} \in_{\delta \alpha \beta \gamma}  j^{\delta}
\end{equation}
we get
\begin{equation}
\label{pomocnicze_28}
(\ast f_{[\alpha \beta} )_{; \gamma]} = \frac{4 \pi}{3c} i \ast j_{\alpha \beta \gamma}
\end{equation}
Then the second part of Eqs. (\ref{Maxwell_equations}) can be equivalently rewritten in the form
\begin{equation}
\label{pomocnicze_29}
f_{[\alpha \beta ; \gamma]}=0
\end{equation}
Adding (\ref{pomocnicze_28}) and (\ref{pomocnicze_29}) we obtain
\begin{equation}
\label{Maxwell_equations_equivalent}
(f_{[\alpha \beta} + \ast f _{[\alpha \beta} )_{; \gamma]} = \frac{4 \pi}{3c} i \ast j_{\alpha \beta \gamma}
\end{equation}
Eqs. (\ref{Maxwell_equations_equivalent}) are equivalent to the Maxwell equations (\ref{Maxwell_equations}).

The spinor images of $f_{\mu \nu}$ are
\begin{eqnarray}
f_{AB} &=& f_{(AB}) = \frac{1}{4} \, S_{AB}^{\ \ \ \mu \nu} f_{\mu \nu}
\\ \nonumber
f_{\dot{A}\dot{B}} &=& f_{(\dot{A}\dot{B})} = \frac{1}{4} S_{\dot{A}\dot{B}}^{\ \ \ \mu \nu} f_{\mu \nu} = \overline{f_{AB}}
\end{eqnarray}
The inverse formula reads
\begin{equation}
f_{\mu \nu} = \frac{1}{2} ( f_{AB} S^{AB}_{\ \ \ \mu \nu} + f_{\dot{A}\dot{B}} S^{\dot{A}\dot{B}}_{\ \ \ \mu \nu} )
\end{equation}
Hence
\begin{equation}
\label{pomocnicze_30}
f_{\mu \nu} + \ast f_{\mu \nu} = f_{AB} S^{AB}_{\ \ \ \mu \nu}
\end{equation}
Inserting (\ref{pomocnicze_30}) into (\ref{Maxwell_equations_equivalent}) one has
\begin{equation}
(f_{AB}S^{AB}_{ \ \ \ [\alpha \beta})_{; \gamma]} = \frac{4 \pi}{3c} i \, \ast j_{\alpha \beta \gamma}
\end{equation}
or
\begin{equation}
\label{pomocnicze_31}
D(f_{AB}S^{AB}) = \frac{4 \pi}{c} i \ast j, \ \ \ast j = \frac{1}{3!} (\ast j_{\alpha \beta \gamma}) \, dx^{\alpha} \wedge dx^{\beta} \wedge dx^{\gamma}
\end{equation}
where $S^{AB}$ is the spinor 2-form (\ref{definicja_SAB}).

In helicity formalism we define
\begin{equation}
f_{a} = \Phi_{a}^{\ AB} f_{AB}, \ f_{\dot{a}} = \Phi_{\dot{a}}^{\ \dot{A} \dot{B}} f_{\dot{A}\dot{B}} = \overline{f_{a}}
\end{equation}
Finally, Eq. (\ref{pomocnicze_31}) in helicity formalism reads
\begin{equation}
\label{pomocnicze_32}
\mathcal{D} (f_{a} \mathcal{S}^{a}) = \frac{4 \pi}{c} i \ast j
\end{equation}
The dotted version of (\ref{pomocnicze_32}) has the form
\begin{equation}
\label{pomocnicze_33}
\mathcal{D} (f_{\dot{a}} \mathcal{S}^{\dot{a}}) = -\frac{4 \pi}{c} i \ast j
\end{equation}
Comparing (\ref{Bianchi_dentities_another_form}) with (\ref{pomocnicze_32}) and (\ref{pomocnicze_33}) we conclude that Eqs. (\ref{Bianchi_dentities_another_form}) can be considered as the gravitational counterpart of the electromagnetic field equations (\ref{pomocnicze_32}) and (\ref{pomocnicze_33}). Moreover, the right hand side of (\ref{Bianchi_dentities_another_form}) plays in gravity the role of the current. 

Within our convention the observer of the 4-velocity $u^{\mu}$ defines the electric field $\mathcal{E}_{j}$ and magnetic field $\mathcal{B}_{j}$ as 
\begin{equation}
\mathcal{E}_{j} = f_{j \mu}  u^{\mu}, \ \ \mathcal{B}_{j} = i \ast f_{j \mu}  u^{\mu}, \ \ j=1,2,3
\end{equation}
Thus for $u^{\mu} = (1,0,0,0)$ we get
\begin{equation}
\mathcal{E}_{j} = f_{j0}, \ \ \mathcal{B}_{j} = i \ast f_{j0}
\end{equation}
These identifications justify the definition (\ref{electric_magnetic_curvature}) of electric and magnetic parts of the free gravitational field.

By (\ref{pomocnicze_30}) we also get
\begin{equation}
\mathcal{E}_{j} - i \mathcal{B}_{j} = f_{j0} + \ast f_{j0} = f_{AB}S^{AB}_{\ \ \ j0}
\end{equation}
Then in an orthonormal basis when (\ref{pomocnicze_25}) holds true one obtains
\begin{equation}
f_{a} = \Phi_{a}^{\ AB}f_{AB} = i \frac{1}{\sqrt{2}} (\mathcal{E}_{a} - i \mathcal{B}_{a}) = i \, \overline{F_{a}}, \ \ a=1,2,3
\end{equation}
where
\begin{equation}
\vec{F} = (F_{1}, F_{2}, F_{3}) = \frac{1}{\sqrt{2}} (\vec{\mathcal{E}} + i \vec{\mathcal{B}})
\end{equation}
is the Riemann-Silberstein vector \cite{b24, b25, b26, b27, b28, b29}.

Finally, we are going to rewrite Eqs. (\ref{pomocnicze_12}) and (\ref{pomocnicze_13}) in terms of the helicity formalism. Multiplying (\ref{pomocnicze_12}) by $\Phi_{a}^{\ AB}\Phi_{b}^{\ CD}$, employing (\ref{zaleznosci_helicity}), (\ref{korespondencja_1}) and (\ref{korespondencja_2}) we get
\begin{eqnarray}
\label{pomocnicze_34}
&& \left( \nabla^{\mu} \nabla_{\mu} + \frac{4 \pi G}{c^{4} } T \right) \mathcal{C}_{ab} + 6 \Phi_{a}^{\ AB} \Phi_{b}^{\ CD} \Phi^{c}_{\ (AB} \Phi^{d}_{\ CD)} \, \mathcal{C}_{ce}\mathcal{C}^{e}_{\ d}
\\ \nonumber
&&= \Phi_{a}^{\ AB} \Phi_{b}^{\ CD} \Phi^{c}_{\ (AB} \Phi^{d}_{\ CD)} \, \frac{2 \pi G}{c^{4}} \, \mathcal{S}_{c}^{\ \varrho \mu} \mathcal{S}_{d}^{\ \sigma \nu} \, \nabla_{\mu} \nabla_{\nu} T_{\varrho \sigma}
\end{eqnarray}
where
\begin{equation}
\nonumber
\mathcal{S}_{c}^{\ \varrho \mu} = \Phi_{c}^{\ AB} S_{AB}^{\ \ \ \varrho \mu}
\end{equation}
Using (\ref{definicja_macierzy_Phi}) and performing straightforward calculations one finds the formula
\begin{equation}
\label{wysumowania_Phi}
\Phi_{a}^{\ AB} \Phi_{b}^{\ CD} \Phi^{c}_{\ (AB} \Phi^{d}_{\ CD)} = \delta^{c}_{(a} \delta^{d}_{b)} - \frac{1}{3} \delta_{ab} \delta^{cd}
\end{equation}
Inserting (\ref{wysumowania_Phi}) into (\ref{pomocnicze_34}) we get
\begin{eqnarray}
&& \left( \nabla^{\mu} \nabla_{\mu} + \frac{4 \pi G}{c^{4} } T \right) \mathcal{C}_{ab} + 6 \left( \mathcal{C}_{ac} \mathcal{C}^{c}_{\ b} - \frac{1}{3} \delta_{ab} \, \mathcal{C}_{cd} \mathcal{C}^{dc}  \right)
\\ \nonumber
&& = \frac{2 \pi G}{c^{4}} \left( \mathcal{S}_{(a}^{\ \ \varrho \mu} \mathcal{S}_{b)}^{\ \ \sigma \nu} - \frac{1}{3}\delta_{ab} \, \mathcal{S}^{c \varrho \mu} \mathcal{S}_{c}^{\ \sigma \nu}  \right) \nabla_{\mu} \nabla_{\nu} T_{\varrho \sigma}
\end{eqnarray}
Employing also the relation \cite{b12}
\begin{equation}
\mathcal{S}^{c \varrho \mu} \mathcal{S}_{c \sigma \nu} = S^{AB \varrho \mu} S_{AB \sigma \nu} = 2 \delta^{\varrho \mu}_{\sigma \nu} - 2i | \det (g_{\alpha \beta}) |^{- \frac{1}{2}} \in^{\varrho \mu \, \bullet \, \bullet}_{\ \ \ \sigma \nu}
\end{equation}
where $ \in^{\varrho \mu \, \bullet \, \bullet}_{\ \ \ \sigma \nu} = \in^{\varrho \mu \varkappa \tau}g_{\sigma \varkappa} g_{\nu \tau}$, and Eq. (\ref{matter_equation}) we obtain
\begin{eqnarray}
\label{pomocnicze_36}
&& \left( \nabla^{\mu} \nabla_{\mu} + \frac{4 \pi G}{c^{4} } T \right) \mathcal{C}_{ab} + 6 \left( \mathcal{C}_{ac} \, \mathcal{C}^{c}_{\ b} - \frac{1}{3} \delta_{ab} \, \mathcal{C}_{cd} \, \mathcal{C}^{dc}  \right)
\\ \nonumber
&& = \frac{2 \pi G}{c^{4}} \left( \mathcal{S}_{(a}^{\ \ \varrho \mu} \mathcal{S}_{b)}^{\ \ \sigma \nu} \, \nabla_{\mu} \nabla_{\nu} T_{\varrho \sigma} - \frac{2}{3} \delta_{ab} \, \nabla^{\mu} \nabla_{\mu} T  \right) 
\end{eqnarray}
The complex conjugation of (\ref{pomocnicze_36}) gives the helicity formulation of (\ref{pomocnicze_13}) as 
\begin{eqnarray}
\label{pomocnicze_37}
&& \left( \nabla^{\mu} \nabla_{\mu} + \frac{4 \pi G}{c^{4} } T \right) \mathcal{C}_{\dot{a}\dot{b}} + 6 \left( \mathcal{C}_{\dot{a}\dot{c}} \, \mathcal{C}^{\dot{c}}_{\ \dot{b}} - \frac{1}{3} \delta_{\dot{a} \dot{b}} \, \mathcal{C}_{\dot{c} \dot{d}} \, \mathcal{C}^{\dot{d}\dot{c}}  \right)
\\ \nonumber
&& = \frac{2 \pi G}{c^{4}} \left( \mathcal{S}_{(\dot{a}}^{\ \ \varrho \mu} \mathcal{S}_{\dot{b})}^{\ \ \sigma \nu} \, \nabla_{\mu} \nabla_{\nu} T_{\varrho \sigma} - \frac{2}{3} \delta_{\dot{a}\dot{b}} \, \nabla^{\mu} \nabla_{\mu} T  \right) 
\end{eqnarray}
It is certainly interesting and important to study further Eqs. (\ref{Bianchi_dentities_another_form}), (\ref{pomocnicze_36}) and (\ref{pomocnicze_37}) in order to write them in a more explicit form which will enable us to compare the helicity approach with the approaches proposed by other authors \cite{b18, b21}.

We leave this problem for the next work since in the present paper we deal mainly with the application of curvature dynamics to the linearized gravity.


\section{Linearized gravity}
\setcounter{equation}{0}

In linearized gravity \cite{b4, b5, b6} we assume that the spacetime metric $g_{\mu \nu}$ has the form of small perturbation of the Minkowski metric $\eta_{\mu \nu} = \textrm{diag} (-1,1,1,1)$
\begin{equation}
g_{\mu \nu} = \eta_{\mu \nu} + h_{\mu \nu}, \ \ | h_{\mu \nu} | \ll 1
\end{equation}
Then the inverse metric reads
\begin{equation}
g^{\mu \nu} = \eta^{\mu \nu} - h^{\mu \nu}, \ \ h^{\mu \nu} = \eta^{\mu \varrho} \eta^{\nu \sigma} \, h_{\varrho \sigma}
\end{equation}
In linearized gravity the lowering and rising of tensorial indices is done by $\eta_{\mu \nu}$ and $\eta^{\mu \nu}$, respectively.

The connection coefficients (\ref{Christofel_symbols}) are now
\begin{equation}
\Gamma^{\mu}_{\ \nu \varrho} = \frac{1}{2} \eta^{\mu \tau} \left( \partial_{\nu} h_{\varrho \tau} + \partial_{\varrho} h_{\nu \tau} - \partial_{\tau} h_{\nu \varrho} \right)
\end{equation}
and the Riemann curvature tensor (\ref{Riemann_curvature_tensor_definition}) reads
\begin{equation}
\label{tensor_Riemanna_liniowy}
R_{\alpha \beta \gamma \delta} = \frac{1}{2} \left( \partial_{\beta} \partial_{\gamma} h_{\alpha \delta} + \partial_{\alpha} \partial_{\delta} h_{\beta \gamma} - \partial_{\alpha} \partial_{\gamma} h_{\beta \delta} - \partial_{\beta} \partial_{\delta} h_{\alpha \gamma} \right)
\end{equation}
Coordinate transformations in linearized gravity are of the form
\begin{equation}
\label{transformacje_zlineryzowanej_grawitacji}
x'^{\alpha} = x^{\alpha} + \xi^{\alpha} , \ \ \alpha = 0,1,2,3
\end{equation}
where $\xi^{\alpha} = \xi^{\alpha} (x^{\beta})$ are small real functions. Under the transformation (\ref{transformacje_zlineryzowanej_grawitacji}) we get
\begin{equation}
\label{transformacja_g_liniowa}
g'_{\mu \nu} = \eta_{\mu \nu} + h'_{\mu \nu}, \ \ h'_{\mu \nu} = h_{\mu \nu} - \partial_{\mu} \xi_{\nu} - \partial_{\nu} \xi_{\mu}
\end{equation}
where $\xi_{\mu} = \eta_{\mu \varrho} \xi^{\varrho}$. Inserting (\ref{transformacja_g_liniowa}) into (\ref{tensor_Riemanna_liniowy}) one quickly concludes that
\begin{equation}
R'_{\alpha \beta \gamma \delta} = R_{\alpha \beta \gamma \delta}
\end{equation}
Therefore the Riemann curvature tensor in linearized gravity is invariant with respect to the transformation (\ref{transformacje_zlineryzowanej_grawitacji}). Since the transformation (\ref{transformacje_zlineryzowanej_grawitacji}) of $h_{\mu \nu}$ has the form of some gauge transformation we will say that the Riemann curvature tensor is gauge invariant. The same concerns also the Ricci tensor, the curvature scalar and the Weyl tensor.

The Ricci tensor in linearized gravity reads
\begin{equation}
\label{Ricci_linearized_gravity}
R_{\alpha \beta} = R^{\delta}_{\ \alpha \beta \delta} = \frac{1}{2} \left\{ \Box h_{\alpha \beta} - \partial_{\alpha} \partial_{\gamma} \left( h^{\gamma}_{\ \beta} - \frac{1}{2} h \, \delta^{\gamma}_{\beta} \right) - \partial_{\beta} \partial_{\gamma} \left( h^\gamma_{\ \alpha} - \frac{1}{2} h \, \delta^{\gamma}_{\alpha} \right) \right\}
\end{equation}
where $\Box = \eta^{\mu \nu} \partial_{\mu} \partial_{\nu} = \partial^{\mu} \partial_{\mu}$ is the d'Alambert operator and $h = \eta^{\mu \nu} h_{\mu \nu}$. Then the curvature scalar $R = R^{\alpha}_{\ \alpha}$ has the form
\begin{equation}
\label{Ricci_skalar_linearized}
R = \Box h - \partial^{\alpha} \partial^{\beta} h_{\alpha \beta}
\end{equation}
Given (\ref{tensor_Riemanna_liniowy}), (\ref{Ricci_linearized_gravity}) and (\ref{Ricci_skalar_linearized}) we can easily find the traceless Ricci tensor $C_{\alpha \beta}$ and the Einstein tensor $G_{\alpha \beta} = R_{\alpha \beta} - \dfrac{1}{2} R \, \eta_{\alpha \beta}$ as functions of $h_{\alpha \beta}$. Bianchi identities (\ref{pomocnicze_3}) and (\ref{pomocnicze_2}) take now the form
\begin{equation}
\label{Bianchi_identities_linearized_1}
\partial_{\alpha} R^{\alpha}_{\ \beta \gamma \delta} = \frac{16 \pi G}{c^{4}} \, \partial_{[\gamma} \widecheck{T}_{\delta ] \beta}
\end{equation}
and 
\begin{equation}
\label{pomocnicze_38}
\partial_{\alpha} C^{\alpha}_{\ \beta \gamma \delta} = \frac{8 \pi G}{c^{4}} \left( \partial_{[\gamma} T_{\delta ] \beta} + \frac{1}{3} \eta_{\beta [ \gamma} \partial_{\delta ]} T \right)
\end{equation}
Eqs. (\ref{drugie_pochodne_Bianchi_identities}) in linearized gravity have much simpler forms
\begin{equation}
\label{dalambercjan_Riemann}
\Box R_{\alpha \beta \gamma \delta} = \frac{16 \pi G}{c^{4}} \left( \partial_{\alpha} \partial_{[\gamma} \widecheck{T}_{\delta ] \beta} - \partial_{\beta} \partial_{[ \gamma} \widecheck{T}_{\delta ] \alpha} \right)
\end{equation}
and by (\ref{Weyl_tensor_definition}), (\ref{Rownania_Einsteina_second_version}) and (\ref{matter_skalar}) the linearized Eqs. (\ref{pomocnicze_16}) read
\begin{eqnarray}
\label{pomocnicze_44}
\Box C_{\alpha \beta \gamma \delta} &=&  \frac{16 \pi G}{c^{4}} \left\{ \partial_{\alpha} \partial_{[\gamma} \widecheck{T}_{\delta ] \beta} - \partial_{\beta} \partial_{[ \gamma} \widecheck{T}_{\delta ] \alpha} \right.
\\ \nonumber
&& \left. \ \ \ \ \ \ \ 
- \frac{1}{2} \Box \left[ (\eta_{\alpha [ \gamma} \widecheck{T}_{\delta ] \beta} - \eta_{\beta [ \gamma} \widecheck{T}_{\delta ] \alpha }) + \frac{1}{3} T \, \eta_{\alpha [ \gamma} \eta_{\delta ] \beta} \right] \right\}
\end{eqnarray}
(compare Eq. (\ref{dalambercjan_Riemann}) with Eq. (44) in \cite{b1}).

In spinorial formalism the main equations (\ref{pomocnicze_7}) and (\ref{pomocnicze_7a}) within the linearized gravity take the form
\begin{equation}
\partial^{D \dot{D}} C_{ABCD} = - \frac{8 \pi G}{c^{4}} \, \partial_{(A |\dot{E}|} T_{BC)}^{\ \ \ \, \dot{D} \dot{E}}
\end{equation}
and
\begin{equation}
\partial^{D \dot{D}} C_{\dot{A} \dot{B} \dot{C} \dot{D}} = - \frac{8 \pi G}{c^{4}} \, \partial_{E ( \dot{A}} T^{DE}_{\ \ \ \dot{B} \dot{C})}
\end{equation}
respectively. Here $\partial^{D \dot{D}} = g^{D \dot{D} \mu} \, \partial_{\mu}$.

Then Eqs. (\ref{pomocnicze_12}) now read
\begin{equation}
\label{dalambercjan_Weyl_undotted}
\Box C_{ABCD} = \frac{2 \pi G}{c^{4}} \, S_{(AB}^{\ \ \ \ \varrho \mu} S_{CD)}^{\ \ \ \ \sigma \nu} \, \partial_{\mu} \partial_{\nu} T_{\varrho \sigma}
\end{equation}
and Eqs. (\ref{pomocnicze_13}) lead to the complex conjugate of (\ref{dalambercjan_Weyl_undotted})
\begin{equation}
\Box C_{\dot{A} \dot{B} \dot{C} \dot{D}} = \frac{2 \pi G}{c^{4}} \, S_{(\dot{A} \dot{B}}^{\ \ \ \ \varrho \mu} S_{\dot{C}\dot{D})}^{\ \ \ \ \sigma \nu} \, \partial_{\mu} \partial_{\nu} T_{\varrho \sigma}
\end{equation}
Then we investigate the helicity formalism in linearized gravity. The field equations (\ref{Bianchi_dentities_another_form}) now take a simpler form
\begin{eqnarray}
d \mathcal{C}^{a}_{\ b} \wedge \mathcal{S}^{b} &=& \frac{8 \pi G}{c^{4}} \left( -d \mathcal{T}^{a}_{\ \ \dot{b}} \wedge \mathcal{S}^{\dot{b}} + \frac{1}{12} \, dT \wedge \mathcal{S}^{a} \right)
\\ \nonumber
d \mathcal{C}^{\dot{a}}_{\ \dot{b}} \wedge \mathcal{S}^{\dot{b}} &=& \frac{8 \pi G}{c^{4}} \left( -d \mathcal{T}_{\; b}^{\ \dot{a}} \wedge \mathcal{S}^{b} + \frac{1}{12} \, dT \wedge \mathcal{S}^{\dot{a}} \right)
\end{eqnarray}
which can be rewritten as
\begin{eqnarray}
\label{helicity_basic_equations}
d \left( \mathcal{C}^{a}_{\ b} - \frac{2 \pi G}{3c^{4}} \, T \, \delta^{a}_{b} \right) \wedge \mathcal{S}^{b} &=& - \frac{8 \pi G}{c^{4}} \, d \mathcal{T}^{a}_{\ \ \dot{b}} \wedge \mathcal{S}^{\dot{b}}
\\ \nonumber
d \left( \mathcal{C}^{\dot{a}}_{\  \dot{b}} - \frac{2 \pi G}{3c^{4}} \, T \, \delta^{\dot{a}}_{\dot{b}} \right) \wedge \mathcal{S}^{\dot{b}} &=&
- \frac{8 \pi G}{c^{4}} \, d \mathcal{T}_{\; b}^{\ \dot{a}} \wedge \mathcal{S}^{b}
\end{eqnarray}
Employing (\ref{definicja_g_mu_AB}), (\ref{definicja_SAB}), (\ref{definicja_macierzy_Phi}), (\ref{definicja_macierzy_Phi_complex_conjugate}) and (\ref{korespondencja_1}) one can find $\mathcal{S}^{a}$ and $\mathcal{S}^{\dot{a}}$. Inserting the results into (\ref{helicity_basic_equations}) and performing straightforward calculations we get the important relations
\begin{eqnarray}
\label{tajemnicze_rownania}
\partial_{0} \mathcal{C}_{ab} &=& i \sum_{c,e} \in_{ace} \partial_{c} \mathcal{C}_{eb} - \frac{8 \pi G}{c^{4}} \left\{ \partial_{0} \left( \mathcal{T}_{b \dot{a}} - \frac{1}{12} \, T \delta_{ba} \right)  \right.
\\ \nonumber
&&  \ \ \ \ \ \ \ \ \ \ \ \ \ \ \ \ \ \ \ \ \ \ \ \ \ \ \ 
\left.  + i \sum_{c,e} \in_{ace} \partial_{c} \left( \mathcal{T}_{b \dot{e}} + \frac{1}{12} \, T \, \delta_{be} \right) \right\}
\\ \nonumber
\sum_{b} \partial_{b} \mathcal{C}_{ab} &=& - \frac{8 \pi G}{c^{4}} \sum_{b} \partial_{b} \left( \mathcal{T}_{a \dot{b}} - \frac{1}{12} \, T \, \delta_{ab} \right)
\end{eqnarray}
and their complex conjugate.

Using (\ref{zaleznosc_miedzy_C_E_B}) and the obvious relations
\begin{equation}
\Re \mathcal{T}_{a \dot{b}} = \frac{1}{2} ( \mathcal{T}_{a \dot{b}} + \mathcal{T}_{b \dot{a}} ), \ \ \Im \mathcal{T}_{a \dot{b}} = \frac{1}{2i} ( \mathcal{T}_{a \dot{b}} - \mathcal{T}_{b \dot{a}} ), \ \ \overline{T}=T
\end{equation}
we equivalently rewrite Eqs. (\ref{tajemnicze_rownania}) as
\begin{eqnarray}
\label{rownania_rozpisane}
\partial_{0} \mathcal{E}_{ab} - \sum_{c,e} \in_{ace} \partial_{c} \mathcal{B}_{eb} &=& \frac{4 \pi G}{c^{4}} \left\{ \partial_{0} \left( \mathcal{T}_{a \dot{b}} + \mathcal{T}_{b \dot{a}} - \frac{1}{6} \, T \delta_{ab} \right) + i \sum_{c,e} \in_{ace} \partial_{c} (\mathcal{T}_{b \dot{e}} - \mathcal{T}_{e \dot{b}} ) \right\} \ \ \ \ \ \ 
\\ \nonumber
\partial_{0} \mathcal{B}_{ab} + \sum_{c,e} \in_{ace} \partial_{c} \mathcal{E}_{eb} &=& \frac{4 \pi G}{c^{4}} \left\{ i \partial_{0} ( \mathcal{T}_{b \dot{a}} - \mathcal{T}_{a \dot{b}} ) - \sum_{c,e} \in_{ace} \partial_{c} \left(  \mathcal{T}_{b \dot{e}} + \mathcal{T}_{e \dot{b}} + \frac{1}{6} \, T \, \delta_{be} \right)  \right\}
\\ \nonumber
\sum_{b} \partial_{b} \mathcal{E}_{ab} &=& \frac{4 \pi G}{c^{4}} \sum_{b} \partial_{b} \left(  \mathcal{T}_{a \dot{b}} + \mathcal{T}_{b \dot{a}} - \frac{1}{6} \, T \, \delta_{ab} \right)
\\ \nonumber
\sum_{b} \partial_{b} \mathcal{B}_{ab} &=& \frac{4 \pi G}{c^{4}} i \sum_{b} \partial_{b} \left(  \mathcal{T}_{a \dot{b}} - \mathcal{T}_{b \dot{a}}  \right)
\end{eqnarray}
If we compare the form of Eqs. (\ref{rownania_rozpisane}) with the form of Maxwell equations it becomes quite clear why $\mathcal{E}_{ab}$ is called the electric part of the free gravitational field and $\mathcal{B}_{ab}$ is the magnetic part of this field. Finally, one can rewrite Eqs. (\ref{pomocnicze_38}) in a similar form to the equations (\ref{rownania_rozpisane}). Simple but tedious manipulations in which the symmetry relations $\mathcal{E}_{ab} = \mathcal{E}_{ba}$, $\mathcal{B}_{ab} = \mathcal{B}_{ba}$, vanishing of the traces $\mathcal{E}^{a}_{\ a}=0$, $\mathcal{B}^{a}_{\ a}=0$ and vanishing of the contraction $C^{a}_{\ bac}=0$ are intensively employed give the following result
\begin{eqnarray}
\label{rownania_rozpisane_2}
\partial_{0} \mathcal{E}_{ab} - \sum_{c,e} \in_{ace} \partial_{c} \mathcal{B}_{eb} &=& \frac{4 \pi G}{c^{4}} \left[ \partial_{b}T_{a0} - \partial_{0} \left( T_{ab} - \frac{1}{3} \, T \, \delta_{ab} \right) \right]
\\ \nonumber
\partial_{0} \mathcal{B}_{ab} + \sum_{c,e} \in_{ace} \partial_{c} \mathcal{E}_{eb} &=&\frac{4 \pi G}{c^{4}} \sum_{c,e} \in_{bce} \partial_{c} \left( T_{ea} - \frac{1}{3} \, T \, \delta_{ea} \right)
\\ \nonumber
\sum_{b} \partial_{b} \mathcal{E}_{ab} &=& \frac{4 \pi G}{c^{4}} \left[ \partial_{a} \left( T_{00} + \frac{1}{3} \, T \right) - \partial_{0} T_{a 0} \right]
\\ \nonumber
\sum_{b} \partial_{b} \mathcal{B}_{ab} &=& \frac{4 \pi G}{c^{4}} \sum_{b,c} \in_{abc} \partial_{b} T_{c0}
\end{eqnarray}
[Note that the constraint equations given by the third and the forth systems in (\ref{rownania_rozpisane}) or (\ref{rownania_rozpisane_2}) can be derived from the first and the second systems of evolution equations of  (\ref{rownania_rozpisane}) or (\ref{rownania_rozpisane_2}), respectively].


\section{Curvature dynamics for gravitational radiation in linearized gravity}
\setcounter{equation}{0}

In this section we deal with the gravitational radiation of a bounded source described by the energy-momentum tensor $T_{\alpha \beta} = T_{\alpha \beta} (\vec{r},t)$ in linearized gravity. Here the curvature dynamics is determined by the equations (\ref{Bianchi_identities_linearized_1}), (\ref{dalambercjan_Riemann}) and the Einstein equations (\ref{Rownania_Einsteina}) which we rewrite in the form
\begin{equation}
\label{Eintein_equations_rewritten}
R_{\alpha \beta} = - \frac{8 \pi G}{c^{4}} \, \widecheck{T}_{\alpha \beta}
\end{equation}
The retarded solution of Eq. (\ref{dalambercjan_Riemann}) reads
\begin{equation}
\label{rozwiazanie_rownania_falowego}
R_{\alpha \beta \gamma \delta} = \frac{4 G}{c^{4}} \int_{\mathbb{R}^{3}} \frac{\left(  
 \partial'_{\beta} \partial'_{[\gamma} \widecheck{T}_{\delta ] \alpha } (\vec{r'}, t') -  \partial'_{\alpha} \partial'_{[\gamma} \widecheck{T}_{\delta ] \beta } (\vec{r'}, t')
\right)_{t'=t- \frac{|\vec{r} - \vec{r'}|}{c} }}{|\vec{r} - \vec{r'}|} d \vec{r'}
\end{equation}
where $\partial'_{\beta} := \dfrac{\partial}{\partial x'^{\beta}}$,... etc., and $\partial'_{0} := \dfrac{1}{c} \dfrac{\partial}{\partial t'}$, $d \vec{r'} = dx'^{1} dx'^{2} dx'^{3}$. It can be easily shown that the integration by parts leads to the following relation
\begin{equation}
\int_{\mathbb{R}^{3}} \frac{\left(  
 \partial'_{\alpha} \partial'_{\beta} \widecheck{T}_{\gamma \delta } (\vec{r'}, t') 
\right)_{t'=t- \frac{|\vec{r} - \vec{r'}|}{c} }}{|\vec{r} - \vec{r'}|} d \vec{r'} 
= \partial_{\alpha} \partial_{\beta} 
\int_{\mathbb{R}^{3}} \frac{\widecheck{T}_{\gamma \delta } \left( \vec{r'}, t-\dfrac{|\vec{r} - \vec{r'}|}{c} \right)}{|\vec{r} - \vec{r'}|}  d \vec{r'}
\end{equation}
Employing this relation in (\ref{rozwiazanie_rownania_falowego}) we get $R_{\alpha \beta \gamma \delta}$ in the form of (\ref{tensor_Riemanna_liniowy}) with 
\begin{equation}
\label{wwzor_na_h_alfabeta}
h_{\alpha \beta} = \frac{4 G}{c^{4}} \int_{\mathbb{R}^{3}} \frac{\widecheck{T}_{\alpha \beta} \left( \vec{r'}, t-\dfrac{|\vec{r} - \vec{r'}|}{c} \right)}{|\vec{r} - \vec{r'}|}  d \vec{r'}
\end{equation}
The functions $h_{\alpha \beta} = h_{\alpha \beta} (\vec{r'}, t)$ are retarded solutions of the equations
\begin{equation}
\label{pomocnicze_5_5}
\Box h_{\alpha \beta} = - \frac{16 \pi G}{c^{4}} \, \widecheck{T}_{\alpha \beta}
\end{equation}
or, equivalently
\begin{equation}
\label{pomocnicze_5_6}
\Box \left( h_{\alpha \beta} - \frac{1}{2} \, h \, \eta_{\alpha \beta} \right) = - \frac{16 \pi G}{c^{4}} \, T_{\alpha \beta}, \ \ h = h^{\alpha}_{\ \alpha}
\end{equation}
Under (\ref{tensor_Riemanna_liniowy}) the Eqs. (\ref{Bianchi_identities_linearized_1}) can be rewritten as
\begin{eqnarray}
\label{pomocnicze_5_7}
&& \partial_{\delta} \left( \Box h_{\beta \gamma} + \frac{16 \pi G}{c^{4}} \, \widecheck{T}_{\beta \gamma} \right) 
- \partial_{\gamma} \left( \Box h_{\beta \delta} + \frac{16 \pi G}{c^{4}} \, \widecheck{T}_{\beta \delta} \right)
\\ \nonumber
&& + \partial_{\beta} \partial_{\gamma} \left[ \partial_{\alpha} \left( h^{\alpha}_{\ \delta} - \frac{1}{2} \, h \, \delta^{\alpha}_{\delta} \right) \right] 
-  \partial_{\beta} \partial_{\delta} \left[ \partial_{\alpha} \left( h^{\alpha}_{\ \gamma} - \frac{1}{2} \, h \, \delta^{\alpha}_{\gamma} \right) \right] =0
\end{eqnarray}
and the Einstein equations (\ref{Eintein_equations_rewritten}) take the form
\begin{equation}
\label{pomocnicze_5_8}
 \left( \Box h_{\beta \delta} + \frac{16 \pi G}{c^{4}} \, \widecheck{T}_{\beta \delta} \right) - \partial_{\beta} \left[ \partial_{\alpha} \left( h^{\alpha}_{\ \delta} - \frac{1}{2} \, h \, \delta^{\alpha}_{\delta} \right)  \right] - \partial_{\delta} \left[  \partial_{\alpha}  \left( h^{\alpha}_{\ \beta} - \frac{1}{2} \, h \, \delta^{\alpha}_{\beta}  \right)  \right] =0
\end{equation}
Employing (\ref{pomocnicze_5_5}) in (\ref{pomocnicze_5_7}) and (\ref{pomocnicze_5_8}) we get
\begin{equation}
\label{pomocnicze_5_9}
\partial_{\beta}  \partial_{\gamma} \left[ \partial_{\alpha}   \left( h^{\alpha}_{\ \delta} - \frac{1}{2} \, h \, \delta^{\alpha}_{\delta}  \right)  \right] - \partial_{\beta}  \partial_{\delta} \left[ \partial_{\alpha}   \left( h^{\alpha}_{\ \gamma} - \frac{1}{2} \, h \, \delta^{\alpha}_{\gamma}  \right)  \right] = 0
\end{equation}
and 
\begin{equation}
\label{pomocnicze_5_10}
\partial_{\beta} \left[ \partial_{\alpha}   \left( h^{\alpha}_{\ \delta} - \frac{1}{2} \, h \, \delta^{\alpha}_{\delta}  \right)  \right] +   \partial_{\delta} \left[ \partial_{\alpha}   \left( h^{\alpha}_{\ \beta} - \frac{1}{2} \, h \, \delta^{\alpha}_{\beta}  \right)  \right] = 0
\end{equation}
for $\beta, \gamma, \delta = 0,1,2,3$.

Changing in (\ref{pomocnicze_5_10}) the index $\beta$ into $\gamma$, then differentiating the equation obtained with respect to $x^{\beta}$ and, finally, comparing the results with Eq. (\ref{pomocnicze_5_9}) one easily concludes that
\begin{eqnarray}
\label{pomocnicze_5_11}
&& \partial_{\gamma} \partial_{\beta} \partial_{\alpha} \left( h^{\alpha}_{\ \delta} - \frac{1}{2} \, h \, \delta^{\alpha}_{\ \delta} \right)=0 
\\ \nonumber
&& \Longrightarrow \ \partial_{\beta} \partial_{\alpha} \left( h^{\alpha}_{\ \delta} - \frac{1}{2} \, h \, \delta^{\alpha}_{\ \delta} \right)= A_{\beta \delta}, \ \beta, \delta = 0,1,2,3
\end{eqnarray}
where $A_{\beta \delta}$ are constants. Since $h_{\alpha \delta}$ tends sufficiently fast to zero at spacial infinity we infer that the constants $A_{\beta \delta}$ are equal to zero
\begin{equation}
A_{\beta \delta} = 0, \ \beta, \delta = 0,1,2,3
\end{equation}
Eq. (\ref{pomocnicze_5_11}) imply the relations
\begin{equation}
\Box \partial_{\alpha} \left( h^{\alpha}_{\ \delta} - \frac{1}{2} \, h \, \delta^{\alpha}_{\ \delta} \right) = 0
\end{equation}
Consequently, by (\ref{pomocnicze_5_6}) one obtains the differential conservation laws 
\begin{equation}
\label{pomocnicze_5_12}
\partial_{\alpha} T^{\alpha}_{\ \delta} = 0
\end{equation}
It can be quickly shown \cite{b4} that employing (\ref{pomocnicze_5_12}) and assuming that the energy-momentum tensor $T_{\alpha \delta}$ tends to zero at spatial infinity fast enough one gets the relation
\begin{equation}
\label{pomocnicze_5_13}
\int_{\mathbb{R}^{3}}  T_{jk}(\vec{r}, t) d \vec{r} = \frac{1}{2c^{2}} \frac{\partial^{2}}{\partial t^{2}} \int_{\mathbb{R}^{3}} T_{00} (\vec{r}, t) x_{j} x_{k} d \vec{r}, \ \ j,k = 1,2,3
\end{equation}
Contraction of (\ref{pomocnicze_5_13}) with $\delta^{jk}$ gives
\begin{equation}
\label{pomocnicze_5_14}
\int_{\mathbb{R}^{3}}  T^{j}_{\ j}(\vec{r}, t) d \vec{r} = \frac{1}{2c^{2}} \frac{\partial^{2}}{\partial t^{2}} \int_{\mathbb{R}^{3}} r^{2} T_{00} (\vec{r}, t)  d \vec{r}, \ \ j,k = 1,2,3
\end{equation}
where $r^{2} = x_{1}^{2} + x_{2}^{2} + x_{3}^{2}$.

To proceed further we will consider the gravitational radiation in vacuum far away from the sources. Thus we assume that 
\begin{equation}
T_{\alpha \beta} (\vec{r},t)=0, \ \ r > r_{0}, \ \ \forall t
\end{equation}
and we consider the points of the configuration 3-space such that 
\begin{equation}
r \gg r_{0}
\end{equation}
Moreover we assume that the velocities of sources are much smaller then the speed of light $c$. Then it can be shown that the perturbation of the metric $h_{\alpha \delta}$ given by (\ref{wwzor_na_h_alfabeta}) can be written in the following form of asymptotic series (see also \cite{b30, b31})
\begin{equation}
h_{\alpha \delta} = \frac{4G}{c^{4}r} \int \widecheck{T}_{\alpha \delta} (\vec{r'}, t - \tfrac{r}{c} ) d \vec{r'} + 
\frac{1}{c^{4}r^{2}} \sum_{k=0}^{\infty} \frac{Y^{(k)}_{\alpha \delta} (t-\frac{r}{c})}{r^{k}} + \frac{1}{c^{5}r} \sum_{j,k=0}^{\infty} \frac{Z^{(j,k)}_{\alpha \delta} (t-\frac{r}{c})}{c^{j} r^{k}}
\end{equation}
Employing the relations (\ref{pomocnicze_5_13}) and (\ref{pomocnicze_5_14}) one gets
\begin{eqnarray}
\nonumber
h_{il} &=& \frac{2G}{3c^{4}r} \ddot{\mathcal{D}}_{il} (t-\tfrac{r}{c}) + \delta_{il} \frac{G}{c^{4} r} \left\{ 2 \int_{\mathbb{R}^{3}} T_{00} (\vec{r'}, t - \tfrac{r}{c} ) d \vec{r'} - \frac{1}{3c^{2}} \int_{\mathbb{R}^{3}} r'^{2}  \ddot{T}_{00} (\vec{r'}, t - \tfrac{r}{c} ) d \vec{r'} \right\} 
\\ 
\label{pomocnicze_5_15}
&& + \frac{1}{c^{4}r^{2}} \sum_{k=0}^{\infty} \frac{Y^{(k)}_{il} (t-\frac{r}{c})}{r^{k}} + \frac{1}{c^{5}r} \sum_{j,k=0}^{\infty} \frac{Z^{(j,k)}_{il} (t-\frac{r}{c})}{c^{j}r^{k}}, \ \ \ i,l=1,2,3
\\ 
\nonumber
h_{i0} &=& \frac{4G}{c^{4}r} \int_{\mathbb{R}^{3}} T_{i0} (\vec{r'}, t- \tfrac{r}{c}) d \vec{r'} + \frac{1}{c^{4}r^{2}} \sum_{k=0}^{\infty} \frac{Y^{(k)}_{i0} (t-\frac{r}{c})}{r^{k}} 
\\ \nonumber
&& + \frac{1}{c^{5}r} \sum_{j,k=0}^{\infty} \frac{Z^{(j,k)}_{i0} (t-\frac{r}{c})}{c^{j}r^{k}}, \ \ \ i=1,2,3
\\ \nonumber
h_{00} &=& \frac{G}{c^{4}r} \left\{ 2 \int_{\mathbb{R}^{3}} T_{00} (\vec{r'}, t - \tfrac{r}{c}) d \vec{r'} + \int_{\mathbb{R}^{3}} r'^{2} \frac{T_{00} (\vec{r'}, t- \frac{r}{c})}{c^{2}} d \vec{r'} \right\}
\\ \nonumber
&& + \frac{1}{c^{4}r^{2}} \sum_{k=0}^{\infty} \frac{Y_{00}^{(k)} (t-\frac{r}{c})}{r^{k}} + \frac{1}{c^{5}r} \sum_{j,k=0}^{\infty} \frac{Z_{00}^{(j,k)} (t- \frac{r}{c})}{c^{j}r^{k}}
\end{eqnarray}
where
\begin{equation}
\mathcal{D}_{il} (t-\tfrac{r}{c}) =\frac{1}{c^{2}} \int_{\mathbb{R}^{3}} (3x'_{i} x'_{l} - \delta_{il} r'^{2}) T_{00} (\vec{r'},  t-\tfrac{r}{c}) d \vec{r'}
\end{equation}
is the quadruple moment of the gravitational sources and the overdot "$\cdot$" stands for the derivative with respect to the time t.

In TT-gauge \cite{b4, b5, b6} we have 
\begin{equation}
\label{form_hTT_in_TT_gauge}
h_{00}^{(TT)}=0, \ \ h_{i0}^{(TT)}=0, \ \ h^{(TT)i}_{\ \ \ \ \ \, i}=0, \ \ \partial_{l} h^{(TT) \ l}_{\ \ \ \ \; i}=0
\end{equation}
Therefore we are left in the TT-gauge with $h^{(TT)}_{il}$, $i,l=1,2,3,$ only and in that gauge we get
\begin{eqnarray}
\label{h_in_TT_gauge}
&& h^{(TT)}_{il} = P_{i}^{\; m} h_{mn} P^{n}_{\ \, l} - \frac{1}{2} P_{il} ( P^{mn} h_{mn})
\\ \nonumber
&& P_{mn} = \delta_{mn} - e_{m} e_{n}, \ \ \ e_{m} = \frac{x_{m}}{r}
\end{eqnarray}
Inserting into (\ref{h_in_TT_gauge}) the first equality of (\ref{pomocnicze_5_15}) one quickly obtains
\begin{eqnarray}
\label{h_il_wgaugu_TT}
h_{il}^{(TT)} &=& \frac{2G}{3c^{4}r} \ddot{\mathcal{D}}_{il}^{(TT)} 
\\ \nonumber
&& + \frac{1}{c^{4}r} \left( P_{i}^{\; m} P^{n}_{\ \, l} - \frac{1}{2} P_{il} P^{mn} \right) \left\{  \frac{1}{r} \sum_{k=0}^{\infty} \frac{Y^{(k)}_{mn} (t-\frac{r}{c})}{r^{k}} + \frac{1}{c} \sum_{j,k=0}^{\infty} \frac{Z_{mn}^{(j,k)} (t-\frac{r}{c})}{c^{j} r^{k}}  \right\}
\end{eqnarray}
where
\begin{equation}
\ddot{\mathcal{D}}_{il}^{(TT)} = P_{i}^{\; m} \ddot{\mathcal{D}}_{mn} (t-\tfrac{r}{c}) P^{n}_{\ \, l} - \frac{1}{2} P_{il} (P^{mn} \ddot{\mathcal{D}}_{mn} (t-\tfrac{r}{c}) )
\end{equation}
Finally, assuming that we consider the point very far away from the sources $r \gg r_{0}$ and assuming also that the velocities of the sources are very small compared to the speed of light $c$ one finds
\begin{equation}
\label{h_il_aproksymacja}
h_{il}^{(TT)} \approx \frac{2G}{3c^{4}r} \left( P_{i}^{\; m} P_{l}^{\; n} - \frac{1}{2} P_{il} P^{mn} \right)
 \ddot{\mathcal{D}}_{mn} (t- \tfrac{r}{c}) = \frac{2G}{3c^{4}r} \ddot{\mathcal{D}}_{il}^{(TT)}
 \end{equation}
 Now we are at the position to consider the electric $\mathcal{E}_{il}$ and magnetic $\mathcal{B}_{il}$ parts of the free gravitational field far away from the sources. Since at such a point $C_{\alpha \beta \gamma \delta} = R_{\alpha \beta \gamma \delta}$ one can rewrite (\ref{electric_magnetic_curvature}) as
\begin{equation}
\mathcal{E}_{il} = R_{0i0l}, \ \mathcal{B}_{il} = i \ast R_{0i0l}, \ i,l=1,2,3
\end{equation}
The first of that equations in linearized gravity reads (see (\ref{tensor_Riemanna_liniowy}))
\begin{equation}
\label{E_il_w_liniowym_przyblizeniu}
\mathcal{E}_{il} = \frac{1}{2} ( \partial_{0} \partial_{i} h_{0l} + \partial_{0} \partial_{l} h_{0i} - \partial_{0} \partial_{0} h_{il} - \partial_{i} \partial_{l} h_{00} )
\end{equation}
The one of advantages of using the curvature variables instead of the metric ones is the independence of the curvature from the gauge. Consequently we can equivalently write (\ref{E_il_w_liniowym_przyblizeniu}) in the TT-gauge defined by (\ref{form_hTT_in_TT_gauge}). Thus we get (compare with \cite{b5})
\begin{equation}
\label{ostateczna_postac_H_il_TT}
\mathcal{E}_{il} = - \frac{1}{2c^{2}} \ddot{h}^{(TT)}_{il}
\end{equation}
Inserting (\ref{h_il_wgaugu_TT}) into (\ref{ostateczna_postac_H_il_TT}) one obtains
\begin{eqnarray}
\mathcal{E}_{il} &=& - \frac{G}{3c^{6}r} \ddddot{\mathcal{D}}^{(TT)}_{il}
\\ \nonumber
&& -\frac{1}{2c^{6}r} \left( P_{i}^{\; m} P_{l}^{\; n} - \frac{1}{2} P_{il} P^{mn} \right)
\left\{ \frac{1}{r} \sum_{k=0}^{\infty} \frac{\ddot{Y}^{(k)}_{mn} (t-\frac{r}{c})}{r^{k}} + \frac{1}{c} \sum_{j,k=0}^{\infty} \frac{\ddot{Z}^{(j,k)}_{mn} (t-\frac{r}{c})}{c^{j}r^{k}} \right\}
\end{eqnarray}
Finally, within the approximation (\ref{h_il_aproksymacja}) we have
\begin{equation}
\label{electric_aproximation}
\mathcal{E}_{il} (\vec{r},t) \approx - \frac{G}{3c^{6}r} \ddddot{\mathcal{D}}^{(TT)}_{il} (\vec{r}, t-\tfrac{r}{c}) = - \frac{G}{3c^{6}r} \left( P_{i}^{\; m} P_{l}^{\; n} - \frac{1}{2} P_{il} P^{mn} \right) \ddddot{\mathcal{D}}_{mn} (t-\tfrac{r}{c})
\end{equation}
and the metric $h_{il}^{(TT)}$ can be expressed as
\begin{equation}
\label{h_wyrazone_przez_E}
h_{il}^{(TT)} (\vec{r},t) = -2c^{2} \int dt \int \mathcal{E}_{il} (\vec{r},t) dt
\end{equation}
The magnetic part $\mathcal{B}_{il}$ reads
\begin{equation}
\label{magnetic_in_TT}
\mathcal{B}_{il} = i \ast R_{0i0l} = -\frac{1}{2} \in_{imn} R^{mn}_{\ \ \ 0l} = \frac{1}{2c} \in_{imn} \frac{\partial \dot{h}^{(TT)\ n}_{\ \ \ \ \; l}}{\partial x_{m}} = \frac{1}{2c} \in_{mn(i} \frac{\partial \dot{h}^{(TT)\ n}_{\ \ \ \ \; l)}}{\partial x_{m}}
\end{equation}
[Remark: To get the last equality in (\ref{magnetic_in_TT}) we used the symmetry of $\mathcal{B}_{il}$, $\mathcal{B}_{il} = \mathcal{B}_{li}$]. 

Substituting (\ref{h_il_wgaugu_TT}) into (\ref{magnetic_in_TT}) one gets
\begin{equation}
\label{magnetic_aproximation}
\mathcal{B}_{il} (\vec{r},t) \approx - \frac{G}{3c^{6}r} \in_{mn(i} \ddddot{\mathcal{D}}^{(TT) \ n}_{\ \ \ \ \; l)} e^{m}
\end{equation}
From (\ref{electric_aproximation}) and (\ref{magnetic_aproximation}) by straightforward calculations we find that 
\begin{eqnarray}
\label{pomocnicze_40}
&& \mathcal{E}_{il} \mathcal{E}^{il} = \mathcal{B}_{il} \mathcal{B}^{il} 
\\ \nonumber
&& \left( \int \mathcal{E}_{il} (\vec{r},t) dt \right) \left( \int \mathcal{E}^{il} (\vec{r},t) dt \right) =
\left( \int \mathcal{B}_{il} (\vec{r},t) dt \right) \left( \int \mathcal{B}^{il} (\vec{r},t) dt \right)=
\\ \nonumber
&& = \frac{1}{4c^{4}} \dot{h}^{(TT)}_{il} (\vec{r},t) \, \dot{h}^{(TT)il} (\vec{r},t) 
\end{eqnarray}
To find the energy flux for the gravitational radiation in linearized gravity we employ the \textsl{Landau - Lifschitz pseudotensor} \cite{b4} $t^{\mu \nu}$
\begin{eqnarray}
\label{pseudotensor_LL}
t^{\mu \nu} &=& -\frac{c^{4}}{16 \pi Gg} \Big\{ \partial_{\varrho} (\sqrt{-g} g^{\mu \nu}) \partial_{\sigma} (\sqrt{-g} g^{\varrho \sigma})  - \partial_{\varrho} (\sqrt{-g} g^{\mu \varrho}) \partial_{\sigma} (\sqrt{-g} g^{\nu \sigma}) 
\\ \nonumber
&& + \frac{1}{2} g^{\mu \nu} g_{\varrho \sigma} \partial_{\delta} (\sqrt{-g} g^{\varrho \tau}) \partial_{\tau} (\sqrt{-g} g^{\delta \sigma)} 
\\ \nonumber
&&- [ g^{\mu \varrho} g_{\sigma \tau} \partial_{\delta} (\sqrt{-g} g^{\nu \tau}) \partial_{\varrho} (\sqrt{-g} g^{\sigma \delta}) 
+ g^{\nu \varrho} g_{\sigma \tau} \partial_{\delta} (\sqrt{-g} g^{\mu \tau}) \partial_{\varrho} (\sqrt{-g} g^{\sigma \delta}) ]
\\ \nonumber
&& + g_{\varrho \sigma} g^{\tau \delta} \partial_{\tau} (\sqrt{-g} g^{\mu \varrho}) \partial_{\delta} (\sqrt{-g} g^{\nu \sigma})
\\ \nonumber
&& + \frac{1}{8} (2 g^{\mu \varrho}g^{\nu \sigma} - g^{\mu \nu} g^{\varrho \sigma}) (2 g_{\tau \delta} g_{\beta \gamma} - g_{\delta \beta} g_{\tau \gamma}) \partial_{\varrho} (\sqrt{-g} g^{\tau \gamma}) \partial_{\sigma} (\sqrt{-g} g^{\delta \beta})  \Big\}
\end{eqnarray}
In the TT-gauge in linearized gravity we get the energy flux vector from (\ref{pseudotensor_LL}) as
\begin{equation}
ct^{i0} = - \frac{c^{4}}{32 \pi G} \frac{\partial h^{(TT)}_{mn}}{\partial x_{i}} \dot{h}^{(TT)mn}, \ \ i=1,2,3
\end{equation}
Employing (\ref{h_il_aproksymacja}) and (\ref{h_wyrazone_przez_E}) one finds
\begin{eqnarray}
\label{pomocnicze_41}
ct^{i0} &=& \frac{G}{72 \pi c^{5}r^{2}} \dddot{\mathcal{D}}^{(TT)}_{mn} \dddot{\mathcal{D}}^{(TT)mn} e^{i} 
\\ \nonumber
&=& \frac{c^{7}}{8 \pi G} \left( \int \mathcal{E}_{mn} (\vec{r},t) dt \right) \left( \int \mathcal{E}^{mn} (\vec{r},t) dt \right) e^{i}
\end{eqnarray}
Using also (\ref{pomocnicze_40}) we can rewrite Eq. (\ref{pomocnicze_41}) in more symmetric form
\begin{eqnarray}
\label{pomocnicze_41_a}
ct^{i0} (\vec{r}, t) &=& \frac{c^{7}}{16 \pi G} \left\{ \int \mathcal{E}_{mn} (\vec{r},t) dt \cdot \int \mathcal{E}^{mn} (\vec{r},t) dt \right.
\\ \nonumber
&& \ \ \ \ \ \ \ \ \ \ 
\left. + \int \mathcal{B}_{mn} (\vec{r},t) dt \cdot \int \mathcal{B}^{mn} (\vec{r},t) dt \right\} e^{i}
\end{eqnarray}
Consequently, the instantaneous power carried by the gravitational radiation in linearized gravity reads
\begin{eqnarray}
\label{pomocnicze_41_b}
L &=&  \int_{r=R_{0}} c t^{i0} (\vec{r}, t) d S_{i}
\\ \nonumber
&=&  \frac{c^{7}}{16 \pi G} \int_{r=R_{0}} \left\{ \int \mathcal{E}_{mn} (\vec{r},t) dt \cdot \int \mathcal{E}^{mn} (\vec{r},t) dt + \int \mathcal{B}_{mn} (\vec{r},t) dt \cdot \int \mathcal{B}^{mn} (\vec{r},t) dt \right\} e^{i} dS_{i}
\end{eqnarray}
where $R_{0} \gg r$. Employing again (\ref{pomocnicze_40}) and then (\ref{electric_aproximation}) one easily gets 
\begin{eqnarray}
\label{postac_L}
L &=&   \frac{c^{7}}{8 \pi G} \int_{r=R_{0}} \left( \int \mathcal{E}_{mn} (\vec{r},t) dt \right) \left( \int \mathcal{E}^{mn} (\vec{r},t) dt \right)  R_{0}^{2} \, d \Omega
\\ \nonumber
&=& \frac{G}{72 \pi c^{5}} \int \bigg\{ \dddot{\mathcal{D}}_{il} (t-\tfrac{R_{0}}{c}) \dddot{\mathcal{D}}^{il} (t-\tfrac{R_{0}}{c}) - 2 \left( e_{l} \dddot{\mathcal{D}}_{i}^{\ l} (t-\tfrac{R_{0}}{c}) \right) \left(e_{m} \dddot{\mathcal{D}}^{im} (t-\tfrac{R_{0}}{c}) \right)  
\\ \nonumber
&& \ \ \ \ \ \ \ \ \ \ \ \ \ \left. +\frac{1}{2} e_{i} e_{l} e_{m} e_{n} \, \dddot{\mathcal{D}}^{il} (t-\tfrac{R_{0}}{c}) \dddot{\mathcal{D}}^{mn} (t-\tfrac{R_{0}}{c}) \right\} d \Omega
\\ \nonumber
&=& \frac{G}{72 \pi c^{5}} \left( 4 \pi - 2 \cdot \frac{4 \pi}{3} + \frac{1}{2} \cdot \frac{4 \pi}{15} \cdot 2 \right) \dddot{\mathcal{D}}_{il} (t-\tfrac{R_{0}}{c}) \dddot{\mathcal{D}}^{il} (t-\tfrac{R_{0}}{c}) 
\\ \nonumber
&=& \frac{G}{45 \pi c^{5}} \, \dddot{\mathcal{D}}_{il} (t-\tfrac{R_{0}}{c}) \dddot{\mathcal{D}}^{il} (t-\tfrac{R_{0}}{c}) 
\end{eqnarray}
Equation (\ref{postac_L}) is of course the well known result in linearized gravity \cite{b4, b5}. Note, that $d \Omega$ in (\ref{postac_L}) stands for the solid angle $d \Omega = \sin \theta \, d \theta d \phi$.


\section{On Bia\l{}ynicki-Birula formula for gravitational energy in linearized gravity}
\setcounter{equation}{0}

I. Bia\l{}ynicki-Birula in his distinguished and inspiring paper \cite{b2} defines the gravitational energy in vacuum in linearized theory as
\begin{equation}
\label{BB_formula}
E_{G} = \frac{c^{4}}{32 \pi^{2} G} \int \frac{\mathcal{E}_{il} (\vec{r}) \mathcal{E}^{il} (\vec{r'}) + \mathcal{B}_{il} (\vec{r}) \mathcal{B}^{il} (\vec{r'}) }{|\vec{r}-\vec{r'}| }d \vec{r} d \vec{r'}
\end{equation}
This formula was further investigated by T. Smo\l{}ka and J. Jezierski \cite{b32} and they have proved that the gravitational energy given by the analogous formula as in (\ref{BB_formula}) but with the factor $\dfrac{c^{4}}{64 \pi^{2}G}$ instead of $\dfrac{c^{4}}{32 \pi^{2}G}$ in the case of localized initial data is equal to the \textsl{Jezierski - Kijowski energy} of linearized gravitational field \cite{b33}. In the present section we show that by taking the factor $\dfrac{c^{4}}{64 \pi^{2}G}$ instead $\dfrac{c^{4}}{32 \pi^{2}G}$ before the integral in (\ref{BB_formula}) one gets the gravitational energy of linearized gravitational radiation in vacuum equal to the energy calculated with the use of Landau-Lifschitz or Einstein energy-momentum pseudotensors.

To this end we first rewrite (\ref{BB_formula}) in the general form
\begin{equation}
\label{BB_formula_witha}
E_{G} = \alpha \int \frac{\mathcal{E}_{il} (\vec{r},t) \mathcal{E}^{il} (\vec{r'},t) + \mathcal{B}_{il} (\vec{r},t) \mathcal{B}^{il} (\vec{r'},t) }{|\vec{r}-\vec{r'}| }d \vec{r} d \vec{r'}
\end{equation}
where $\alpha$ is some positive parameter which is to be found. 

From (\ref{electric_magnetic_curvature}), (\ref{pomocnicze_44}) and (\ref{rownania_rozpisane_2}) one infers that in vacuum the following equations are fulfilled 
\begin{equation}
\partial_{l} \mathcal{E}^{il} = 0, \ \ \partial_{l} \mathcal{B}^{il} = 0
\end{equation}
and
\begin{equation}
\label{vacuum_equations_for_EB}
\Box \mathcal{E}_{il} = 0 , \ \ \Box \mathcal{B}_{il} = 0 
\end{equation}
We consider the solutions of Eqs. (\ref{vacuum_equations_for_EB}) as the superpositions of the plane wave solutions
\begin{eqnarray}
\label{definicja_E_vacuum}
\mathcal{E}_{il} (\vec{r},t) &=& \int \frac{d \vec{k}}{(2 \pi)^{3}} \left\{ \widetilde{\mathcal{E}}^{-}_{il} ( \vec{k} ) \exp \{ i (\vec{k} \cdot \vec{r} - \omega t) \} + \widetilde{\mathcal{E}}_{il}^{+} (\vec{k}) \exp \{- i (\vec{k} \cdot \vec{r} - \omega t) \} \right\}
\\ \nonumber
&=& \int \frac{d \vec{k}}{(2 \pi)^{3}} \widetilde{\mathcal{E}}_{il} (\vec{k}, t) \exp \{ i \, \vec{k} \cdot \vec{r} \}
\end{eqnarray}
and
\begin{eqnarray}
\label{definicja_B_vacuum}
\mathcal{B}_{il} (\vec{r},t) &=& \int \frac{d \vec{k}}{(2 \pi)^{3}} \left\{ \widetilde{\mathcal{B}}^{-}_{il} ( \vec{k} ) \exp \{ i (\vec{k} \cdot \vec{r} - \omega t) \} + \widetilde{\mathcal{B}}_{il}^{+} (\vec{k}) \exp \{- i (\vec{k} \cdot \vec{r} - \omega t) \} \right\}
\\ \nonumber
&=& \int \frac{d \vec{k}}{(2 \pi)^{3}} \widetilde{\mathcal{B}}_{il} (\vec{k}, t) \exp \{ i \, \vec{k} \cdot \vec{r} \}
\end{eqnarray}
where
\begin{eqnarray}
\nonumber
\widetilde{\mathcal{E}}_{il} (\vec{k}, t) &=& \widetilde{\mathcal{E}}_{il}^{-} (\vec{k}) \exp \{ -i \omega t \} +
     \widetilde{\mathcal{E}}_{il}^{+} (-\vec{k}) \exp \{ i \omega t \}
\\ \nonumber
\widetilde{\mathcal{B}}_{il} (\vec{k}, t) &=& \widetilde{\mathcal{B}}_{il}^{-} (\vec{k}) \exp \{ -i \omega t \} +
     \widetilde{\mathcal{B}}_{il}^{+} (-\vec{k}) \exp \{ i \omega t \}
\end{eqnarray}
and $\omega = ck$, $k=| \vec{k} |$.

From the fact that $\mathcal{E}_{il} (\vec{r},t)$ and $\mathcal{B}_{il} (\vec{r},t)$ are real functions one quickly gets the relations
\begin{equation}
\label{wlasnosci_sprzezen}
\widetilde{\mathcal{E}}^{+}_{il} (\vec{k}) = \overline{\widetilde{\mathcal{E}}^{-}_{il} (\vec{k})}, \ \ 
\widetilde{\mathcal{E}}_{il} (\vec{k},t) = \overline{\widetilde{\mathcal{E}}_{il} (-\vec{k},t)}, \ \ 
\widetilde{\mathcal{B}}^{+}_{il} (\vec{k}) = \overline{\widetilde{\mathcal{B}}^{-}_{il} (\vec{k})}, \ \ 
\widetilde{\mathcal{B}}_{il} (\vec{k},t) = \overline{\widetilde{\mathcal{B}}_{il} (-\vec{k},t)}
\end{equation}
Inserting (\ref{definicja_E_vacuum}) and (\ref{definicja_B_vacuum}) into (\ref{BB_formula_witha}), performing the respective integration and employing also (\ref{wlasnosci_sprzezen}) we obtain
\begin{equation}
\label{another_proposition_forE}
E = 4 \pi \alpha  \int \frac{d \vec{k}}{(2\pi )^{3} k^{2}} \left( \widetilde{\mathcal{E}}_{il} (\vec{k},t) \overline{\widetilde{\mathcal{E}}^{il} (\vec{k},t)} + \widetilde{\mathcal{B}}_{il} (\vec{k},t) \overline{\widetilde{\mathcal{B}}^{il} (\vec{k},t)}  \right)
\end{equation}
According to (\ref{pomocnicze_5_5}) the perturbation of the metric $h_{\mu \nu}$ in vacuum fulfills the wave d'Alambert equation. Therefore we represent $h_{\mu \nu} (\vec{r},t)$ as a superposition of the plane wave solutions
\begin{eqnarray}
\label{propozycja_h_vacuum}
h_{\mu \nu} (\vec{r},t) &=& \int \frac{d \vec{k}}{(2 \pi)^{3}} \left\{ \widetilde{h}^{-}_{\mu \nu} ( \vec{k} ) \exp \{ i (\vec{k} \cdot \vec{r} - \omega t) \} + \widetilde{h}_{\mu \nu}^{+} (\vec{k}) \exp \{- i (\vec{k} \cdot \vec{r} - \omega t) \} \right\}
\\ \nonumber
&=& \int \frac{d \vec{k}}{(2 \pi)^{3}} \widetilde{h}_{\mu \nu} (\vec{k}, t) \exp \{ i \, \vec{k} \cdot \vec{r} \}
\end{eqnarray}
where
\begin{equation}
\nonumber
\widetilde{h}_{\mu \nu} (\vec{k}, t) := \widetilde{h}_{\mu \nu}^{-} (\vec{k}) \exp \{ -i \omega t \} +
     \widetilde{h}_{\mu \nu}^{+} (-\vec{k}) \exp \{ i \omega t \}
\end{equation}
Since all the functions $h_{\mu \nu} (\vec{r},t)$ are real we find the relations analogous to (\ref{wlasnosci_sprzezen})
\begin{equation}
\label{wlasnosci_h_falka}
\widetilde{h}_{\mu \nu}^{+} (\vec{k}) = \overline{ \widetilde{h}_{\mu \nu}^{-} (\vec{k}) }, \ \ 
\widetilde{h}_{\mu \nu} (\vec{k},t) = \overline{ \widetilde{h}_{\mu \nu} (-\vec{k},t) }
\end{equation}
For gravitational field in vacuum one can choose the \textsl{transverse-traceless gauge} (the \textsl{TT-gauge}) defined by (\ref{form_hTT_in_TT_gauge}) \cite{b5}. In that gauge the relation (\ref{ostateczna_postac_H_il_TT}) holds true and substituting (\ref{definicja_E_vacuum}) and (\ref{propozycja_h_vacuum}) into (\ref{ostateczna_postac_H_il_TT}) we get
\begin{equation}
\label{Postac_E_proznia}
\widetilde{\mathcal{E}}_{il} (\vec{k},t) = \frac{\omega^{2}}{2 c^{2}} \, \widetilde{h}_{il}^{(TT)} (\vec{k},t)
\end{equation}
Analogously, inserting (\ref{definicja_B_vacuum}) and (\ref{propozycja_h_vacuum}) into (\ref{magnetic_in_TT}) one obtains
\begin{eqnarray}
\label{Postac_B_proznia}
\mathcal{B}_{il} (\vec{k}, t) &=& - \frac{\omega}{2c} \Big\{  \in_{mn(i} \widetilde{h}^{(TT)-}_{l)m} (\vec{k}) \cdot k_{n} \exp \{ -i \omega t \} 
\\ \nonumber
&& \ \ \ \ \ \ \ \ -  \in_{mn(i} \widetilde{h}^{(TT)+}_{l)m} (-\vec{k}) \cdot k_{n} \exp \{ i \omega t \} \Big\}
\end{eqnarray}
Substituting (\ref{Postac_E_proznia}) and (\ref{Postac_B_proznia}) into (\ref{another_proposition_forE}), using also the fact that Eqs. (\ref{form_hTT_in_TT_gauge}) and (\ref{propozycja_h_vacuum}) imply the relations
\begin{equation}
\label{transwersalnosc_ha}
k^{l} \, \widetilde{h}^{(TT)-}_{il} (\vec{k}) = 0, \ \ k^{l} \, \widetilde{h}^{(TT)+}_{il} (\vec{k})=0
\end{equation}
we obtain
\begin{eqnarray}
\label{postac_E_2}
E &=& 2 \pi \alpha \int \frac{d \vec{k}}{(2 \pi)^{3}} \frac{\omega^{2}}{c^{2}} \left\{ \widetilde{h}^{(TT)-}_{il} (\vec{k}) \cdot \overline{ \widetilde{h}^{(TT)-il} (\vec{k}) } +  \widetilde{h}^{(TT)+}_{il} (\vec{k}) \cdot \overline{ \widetilde{h}^{(TT)+il} (\vec{k}) } \right\}
\\ \nonumber
&=& 4 \pi \alpha \int \frac{d \vec{k}}{(2 \pi)^{3}} \frac{\omega^{2}}{c^{2}} \, \widetilde{h}^{(TT)-}_{il} (\vec{k}) \cdot \overline{ \widetilde{h}^{(TT)-il} (\vec{k}) }
\end{eqnarray}
From (\ref{postac_E_2}) one quickly concludes that the B-B gravitational energy is independent of time. 

Now we compare the result (\ref{postac_E_2}) with the gravitational energy in linearized gravity as calculated with the use of the Landau-Lifschitz pseudotensor $t^{\mu \nu}$ given by (\ref{pseudotensor_LL}). Using in (\ref{pseudotensor_LL}) the TT-gauge and performing simple manipulations one finds
\begin{eqnarray}
\label{t_00_uproszczone}
t^{00} &=& \frac{c^{4}}{16 \pi G} \left\{ \frac{1}{2} \eta^{00} \, \partial_{\delta} h^{(TT) \ \tau}_{\ \ \ \ \; \sigma} \cdot \partial_{\tau} h^{(TT) \delta \sigma} + \frac{1}{2c^{2}} \, \dot{h}^{(TT)}_{\delta \beta} \cdot  \dot{h}^{(TT)\delta \beta} \right.
\\ \nonumber
&& \ \ \ \ \ \ \ \ \ \ \left.  - \frac{1}{4} \eta^{00} \, \partial_{\varrho} h^{(TT)}_{\delta \beta} \cdot \partial^{\varrho} h^{(TT)\delta \beta} \right\}
\\ 
\nonumber
&=& \frac{c^{4}}{16 \pi G} \left\{ \frac{1}{4c^{2}} \left( \dot{h}^{(TT)}_{il} \cdot \dot{h}^{(TT)il}-\ddot{h}^{(TT)}_{il} \cdot h^{(TT)il} \right) \right.
\\ \nonumber
&& \ \ \ \ \ \ \ \ \ \ \left. + \frac{1}{4} \partial^{l} \left( \partial_{l} h^{(TT)}_{im} \cdot h^{(TT)im}   -  \partial_{m} h^{(TT)}_{il} \cdot h^{(TT)im}   \right)  \right\}
\end{eqnarray}
Neglecting in (\ref{t_00_uproszczone}) the last term being a space divergence we get the gravitational energy in linearized gravity within the Landau-Lifschitz formalisms as
\begin{equation}
\label{postac_E_LL}
E_{(LL)} = \frac{c^2}{64 \pi G} \int \left\{ \dot{h}^{(TT)}_{il} (\vec{r},t) \cdot \dot{h}^{(TT)il} (\vec{r}, t) - \ddot{h}^{(TT)}_{il} (\vec{r},t) \cdot h^{(TT)il} (\vec{r},t)   \right\} d \vec{r}
\end{equation}
Inserting into (\ref{postac_E_LL}) the Fourier expansion (\ref{propozycja_h_vacuum}) and employing (\ref{wlasnosci_h_falka}) one finally gets
\begin{eqnarray}
\label{postac_E_LL_ostateczna}
E_{(LL)} &=& \frac{c^{4}}{32 \pi G} \int \frac{d \vec{k}}{(2 \pi)^{3}} \frac{\omega^{2}}{c^{2}} \left\{ \widetilde{h}^{(TT)-}_{il} (\vec{k}) \cdot \overline{\widetilde{h}^{(TT)-il} (\vec{k})} + \widetilde{h}^{(TT)+}_{il} (\vec{k}) \cdot \overline{\widetilde{h}^{(TT)+il} (\vec{k})}  \right\}
\\ \nonumber
&=& \frac{c^{4}}{16 \pi G} \int \frac{d \vec{k}}{(2 \pi)^{3}} \frac{\omega^{2}}{c^{2}} \, \widetilde{h}^{(TT)-}_{il} (\vec{k}) \cdot \overline{\widetilde{h}^{(TT)-il} (\vec{k})}  
\end{eqnarray}
Comparing this with (\ref{postac_E_2}) we conclude that 
\begin{equation}
\label{dowod_na_alpha}
E=E_{(LL)} \ \Longleftrightarrow \ \alpha = \frac{c^{4}}{64 \pi^{2} G}
\end{equation}
This is exactly what we wanted to prove.

Analogously one can show that the gravitational energy of the linearized gravitational field in vacuum calculated with the use of the \textsl{Einstein energy-momentum pseudotensor} \cite{b4, b34} is equal to $E_{(LL)}$ given by (\ref{postac_E_LL_ostateczna}) and also to $E$ defined by (\ref{BB_formula_witha}) for $\alpha$ given by (\ref{dowod_na_alpha}).

[This last statement is in the full agreement with the results of \cite{b35} for a weak gravitational wave which prove that in this case the Einstein, Landau-Lifschitz and other energy-momentum pseudotensors in the TT-gauge give identical results for the density of gravitational energy].

Motivated by the Bia\l{}ynicki-Birula formula for the gravitational energy and by the formula for the \textsl{super-Poynting vector} given by Eq. (41) in \cite{b21}) we propose the following formula for the momentum of linearized gravitational field
\begin{equation}
\label{momentum_0}
P^{i} = \alpha' \int \frac{\in^{ilm} \mathcal{E}_{l}^{\; n} (\vec{r},t) \, \mathcal{B}_{mn} (\vec{r'},t)}{|\vec{r}-\vec{r'}|}  d \vec{r'} d \vec{r}
\end{equation}
where $\alpha'$ is a parameter which is to be chosen so that
\begin{equation}
\label{momentum_1}
P^{i} = P^{i}_{(LL)} = \frac{1}{c} \int (-g) t^{i0} (\vec{r},t) d \vec{r}
\end{equation}
where $P^{i}_{(LL)}$ is the momentum of linearized gravitational field calculated for the Landau-Lifschitz pseudotensor (\ref{pseudotensor_LL}) in the TT-gauge. Using the TT-gauge (\ref{form_hTT_in_TT_gauge}) in (\ref{momentum_1}) with (\ref{pseudotensor_LL}) one gets
\begin{eqnarray}
\label{momentum_2}
P^{i}_{(LL)} &=& \frac{c^{2}}{32 \pi G} \int \left\{ -\dot{h}^{(TT)}_{mn} (\vec{r},t) \partial^{i} h^{(TT)mn} (\vec{r},t)   \right.
\\ \nonumber
&& \ \ \ \ \ \ \ \ \ \ \ \ \  \left.
+ \partial_{m}  \left( 2 h^{(TT)i}_{\ \ \ \ \ \; l} (\vec{r},t)  \cdot \dot{h}^{(TT)lm} (\vec{r},t)  \right)  \right\}  d \vec{r}
\end{eqnarray}
Neglecting the last term of (\ref{momentum_2}) being a space divergence, inserting (\ref{propozycja_h_vacuum}) into (\ref{momentum_2}) and employing also (\ref{wlasnosci_h_falka}) we obtain
\begin{eqnarray}
P^{i}_{(LL)} &=& \frac{c^{3}}{32 \pi G} \int \frac{d \vec{k}}{(2 \pi)^{3}} k^{i} \frac{\omega}{c}  \left\{ \widetilde{h}^{(TT)-}_{mn} (\vec{k}) \cdot \overline{\widetilde{h}^{(TT)-mn} (\vec{k})}  +  \widetilde{h}^{(TT)+}_{mn} (\vec{k}) \cdot \overline{\widetilde{h}^{(TT)+mn} (\vec{k})}   \right\} \ \ \ \ 
\\ \nonumber
&=& \frac{c^{3}}{16 \pi G} \int \frac{d \vec{k}}{(2 \pi)^{3}} k^{i} \frac{\omega}{c}  \left\{ \widetilde{h}^{(TT)-}_{mn} (\vec{k}) \cdot \overline{\widetilde{h}^{(TT)-mn} (\vec{k})}    \right\}
\end{eqnarray}
Then substituting (\ref{definicja_E_vacuum}) and (\ref{definicja_B_vacuum}) into (\ref{momentum_0}), performing the integration and using also (\ref{wlasnosci_sprzezen}) one gets
\begin{equation}
\label{momentum_3}
P^{i} = 4 \pi \alpha' \int \frac{d \vec{k}}{(2 \pi)^{3} k^{2}} \in^{ilm} \widetilde{\mathcal{E}}_{ln} (\vec{k},t) \cdot \overline{\widetilde{\mathcal{B}}_{m}^{\ \; n} (\vec{k},t)}
\end{equation}
Employing (\ref{Postac_E_proznia}) and (\ref{Postac_B_proznia}) in (\ref{momentum_3}) and using also (\ref{wlasnosci_h_falka}) and (\ref{transwersalnosc_ha}) we finally find
\begin{eqnarray}
\label{momentum_4}
P^{i} &= & \pi \alpha' \int \frac{d \vec{k}}{(2 \pi)^{3}} k^{i} \frac{\omega}{c} 
\left\{ \widetilde{h}^{(TT)-}_{mn} (\vec{k}) \cdot \overline{ \widetilde{h}^{(TT)-mn} (\vec{k}) } +  \widetilde{h}^{(TT)+}_{mn} (\vec{k}) \cdot \overline{ \widetilde{h}^{(TT)+mn} (\vec{k}) } \right\}
\\ \nonumber
&=& 2 \pi \alpha' \int \frac{d \vec{k}}{(2 \pi)^{3}} k^{i} \frac{\omega}{c} \, 
 \widetilde{h}^{(TT)-}_{mn} (\vec{k}) \cdot \overline{ \widetilde{h}^{(TT)-mn} (\vec{k}) }  
\end{eqnarray}
Thus one quickly concludes that the equality (\ref{momentum_1}) in linearized gravity holds true iff
\begin{equation}
\alpha' = \frac{c^{3}}{32 \pi^{2}G}
\end{equation}

\section{Concluding remarks}

In this work we have presented the curvature and conformal curvature dynamics developed by J.W. van Holten \cite{b1} in the spinorial and then helicity formalisms. In our opinion especially the helicity formalism in the form considered by Pleba\'nski \cite{b3} provides one with a useful and natural language since that formalism the electric and magnetic parts arise in a natural way as the real and imaginary, respectively, parts of the "helicity image" of the Weyl tensor (see Eq. (\ref{zaleznosc_miedzy_C_E_B}) of this paper), and the transformation rules founded on the structure group $SO(3; \mathbb{C})$ are directly included. Therefore of a great interest should be a comparison of the "helicity approach" with other approaches considered for example in \cite{b18, b21}. We are going to deal with this issue in the next work. In the current work we apply the general results to the much simpler problem i.e. to the linearized gravitational radiation.

\end{document}